\documentclass[twocolumn]{aastex7}
\usepackage{amsmath}
\usepackage{gensymb}
\usepackage[normalem]{ulem}
\usepackage{subfigure}
\usepackage{booktabs}
\usepackage{color}
\usepackage{graphicx}
\usepackage{epstopdf}
\usepackage{xcolor}
\usepackage{appendix}
\usepackage{rotating}
\usepackage{CJKutf8}
\usepackage[table]{xcolor}
\usepackage{placeins} 
\usepackage{multirow}
\usepackage{soul}

\shorttitle{Detect CSST Slitless Spectral Trace with YOLO}
\shortauthors{Zhou et al.}

\begin{document}

\title{CSST Slitless Spectra: Target Detection and Classification with YOLO}

\author[0000-0002-8030-4383]{Yingying Zhou}
\affiliation{Key Laboratory of Space Astronomy and Technology, 
National Astronomical Observatories, 
Chinese Academy of Sciences, 
Beijing 100101, P.R. China}
\email{zhouyingying@bao.ac.cn}

\author[0000-0002-1802-6917]{Chao Liu}
\affiliation{Key Laboratory of Space Astronomy and Technology, 
National Astronomical Observatories, 
Chinese Academy of Sciences, 
Beijing 100101, P.R. China}
\affiliation{School of Astronomy and Space Science, University of Chinese Academy of Sciences, 19A Yuquan Road, Shijingshan District, Beijing 100049, P.R. China}
\email[show]{liuchao@nao.cas.cn}

\author[0000-0003-3347-7596]{Hao Tian}
\affiliation{Key Laboratory of Space Astronomy and Technology, 
National Astronomical Observatories, 
Chinese Academy of Sciences, 
Beijing 100101, P.R. China}
\email{tianhao@nao.cas.cn}

\author[0000-0001-7314-4169]{Xin Zhang}
\affiliation{Key Laboratory of Space Astronomy and Technology, 
National Astronomical Observatories, 
Chinese Academy of Sciences, 
Beijing 100101, P.R. China}
\email{zhangx@bao.ac.cn}

\author[0000-0001-6800-7389]{Nan Li}
\affiliation{Key Laboratory of Space Astronomy and Technology, 
National Astronomical Observatories, 
Chinese Academy of Sciences, 
Beijing 100101, P.R. China}
\email{nan.li@nao.cas.cn}

\begin{abstract}
	
Addressing the spatial uncertainty and spectral blending challenges in CSST slitless spectroscopy, we present a deep learning-driven, end-to-end framework based on the You Only Look Once (YOLO) models. This approach directly detects, classifies, and analyzes spectral traces from raw 2D images, bypassing traditional, error-accumulating pipelines. YOLOv5 effectively detects both compact zero-order and extended first-order traces even in highly crowded fields. Building on this, YOLO11 integrates source classification (star/galaxy) and discrete astrophysical parameter estimation (e.g., redshift bins), showcasing complete spectral trace analysis without other manual preprocessing. Our framework processes large images rapidly, learning spectral-spatial features holistically to minimize errors. We achieve high trace detection precision (YOLOv5) and demonstrate successful quasar identification and binned redshift estimation (YOLO11). This study establishes machine learning as a paradigm shift in slitless spectroscopy, unifying detection, classification, and preliminary parameter estimation in a scalable system. Future research will concentrate on direct, continuous prediction of astrophysical parameters from raw spectral traces.
\end{abstract}


\section{Introduction} \label{sec:intro}

\subsection{Overview of CSST and Motivation}

The China Space Station Telescope (CSST) is set to serve as China’s premier astronomical facility over the coming decades, with a projected operational lifespan of ten years. Designed to survey approximately 17,500 square degrees of the sky, CSST combines high spatial resolution ($\leq$ 0.15 arcseconds) with a multi-channel observational framework. Its imaging capabilities span seven photometric bands (NUV, u, g, r, i, z, y), achieving a detection depth of 26 mag in the g band with 300-second exposures and an average depth of 25.5 mag in NUV, u, r, i, and z bands. Complementing these imaging surveys, CSST’s slitless spectroscopy covers three wavelength ranges: GU (255–420 nm), GV (400–650 nm), and GI (620–1000 nm), with a spectral resolution of R $\lesssim$ 200 and broadband sensitivities reaching 22–23 mag. Ultra-deep surveys over 400 square degrees will extend these limits by an additional magnitude, enabling unprecedented studies of faint astronomical objects \citep{zhancsst2011,zhancsst2018,zhancsst2021}.

CSST holds exceptional potential for resolving the Milky Way’s central bulge, where it will obtain low-resolution spectra for 10–20 million stars and multi-band photometry for 1–2 billion stars. These data will refine measurements of stellar metallicities, kinematics, and ages, offering new constraints on the properties of past merger events (e.g., progenitor mass, merger chronology) and identifying previously undetected accretion remnants. Such insights promise to advance reconstructions of the Milky Way’s hierarchical assembly history.

Beyond the Galaxy, CSST’s wide-field capabilities will resolve individual stars in nearby galaxies, enabling detailed analyses of their star formation histories, metallicity distributions, and tidal interactions. By mapping stellar streams and halo substructures, the telescope will shed light on galaxy evolution mechanisms, including satellite accretion, dark matter halo dynamics, and the role of mergers in shaping disk galaxies.

CSST’s deep, wide-area surveys will probe cosmological phenomena such as dark energy, large-scale structure growth, and galaxy clustering. Its photometric redshift accuracy and slitless spectroscopy will enhance measurements of weak gravitational lensing, baryon acoustic oscillations (BAO), and redshift-space distortions. These datasets will tighten constraints on cosmological parameters, including the Hubble constant and the equation of state of dark energy, while ultra-deep fields will trace early galaxy formation, bridging observations of the high-redshift universe with local galactic ecosystems.

By unifying studies of the Milky Way, nearby galaxies, and the distant universe, CSST will establish a cross-scale framework for addressing fundamental questions in astrophysics, from the assembly of stellar halos to the expansion history of the cosmos.

CSST employs an innovative focal plane configuration distinct from conventional telescope designs. To maximize observational throughput, its primary focal plane integrates both slitless spectroscopic detectors and multi-band imaging sensors simultaneously. While this dual-mode design enhances efficiency, it introduces a unique challenge: during initial observational phases, slitless spectroscopic data lack contemporaneous multi-color photometric references. Without overlapping imaging to anchor positional coordinates, the spatial origins of slitless spectral signals remain ambiguous. Consequently, the absence of direct source localization prevents precise extraction of spectra.

CSST will produce unprecedented data volumes, presenting a daunting computational and analytical challenge. A critical issue arises in densely populated regions, such as the Milky Way’s core, where slitless spectroscopic observations inherently produce overlapping spectral traces from adjacent sources. In crowded stellar environments like the Galactic bulge, the absence of physical slits allows light from neighboring stars to blend, creating composite signals that complicate spectral disentanglement and source-specific analysis. Legacy pipelines, optimized for isolated sources, often exclude overlapping spectral regions to simplify analysis, inadvertently sacrificing critical astrophysical information. Retaining and interpreting these overlaps, however, is vital for reconstructing accurate stellar properties (e.g., chemical abundances, kinematic properties) and ensuring the scientific fidelity of CSST’s outputs.

To address the dual challenges of (1) spatial uncertainties arising from non-simultaneous imaging and (2) spectral blending in crowded stellar fields, we present a deep learning-driven, end-to-end framework that decodes stellar properties directly from two-dimensional (2D) slitless spectral traces. By tackling these limitations, our methodology aims to enhance the fidelity of future CSST data products, enabling breakthroughs in stellar astrophysics, galactic archaeology, and cosmological studies.

Extracting astrophysical parameters from 2D spectral images is achievable because these images preserve the full wavelength-flux relationship along the dispersion axis—mirroring the diagnostic power of 1D spectra—while simultaneously encoding spatial information along the cross-dispersion axis. For point sources (e.g., stars), the spatial axis captures the telescope’s point spread function (PSF), enabling precise modeling of how light from individual stars spreads across the detector, which helps disentangle overlapping spectra in crowded fields. For extended sources (e.g., galaxies), the spatial axis reveals spatial gradients like velocity or metallicity,  providing multidimensional constraints unavailable in 1D data. With both spectral and spatial information, advanced computational techniques (e.g., machine learning, Bayesian inference) could deblend overlapping signals, resolve parameter degeneracies, and correlate spectral-spatial patterns with physical properties.   

This dual-axis synergy\footnote{It means combining spectral and position information to solve problems that one alone can’t.} positions 2D spectral analysis as an indispensable tool for surveys like CSST, where the extreme stellar density of the Galactic bulge and the structural complexity of galaxies demand multidimensional diagnostics to overcome the limitations of traditional 1D methods.

In this study, we focus on the framework’s foundational phase: automated detection of spectral traces in unprocessed slitless images and precise delineation of bounding boxes to isolate individual sources. This approach mitigates positional ambiguity in early-phase CSST data—where concurrent multi-band imaging is unavailable—facilitating subsequent spectral extraction and parameter derivation. Critically, we validate the model’s proficiency in identifying sources within ultra-dense regions, underscoring its potential to maximize CSST’s discovery potential in environments where traditional methods falter.

\subsection{The characteristic and challenges of the our dataset}

As previously noted, CSST data processing faces two critical challenges: (1) positional ambiguity (due to the absence of concurrent imaging for spatial anchoring) and (2) spectral blending (in crowded stellar fields). To address the first challenge, object detection techniques from computer vision—such as bounding box localization—could be adapted to identify and spatially pinpoint spectral traces directly in 2D slitless images, bypassing the need for prior imaging. 

Object detection aims to identify specific semantic objects within digital images and videos. These objects can belong to various categories, such as humans, buildings, or cars. Different form items commonly seen in daily life, we focus on different orders of sltiless spectra, which possess unique characteristics and leads to some specific challenges. 

\subsubsection{Small Target Challenge}
Our analysis focuses on a class of compact targets—the zero-order spectral image—for stars it has a size of $\sim$8$\times$8 pixels. these small targets embedded within ultra-high-resolution image frames (9216$\times$9232 pixels). Detecting them poses unique computational and methodological challenges: high-resolution data strains memory and processing capacity, while tiny targets easily obscured by noise, background variability, or false alarms—demand specialized detection strategies. The simplest way to solve the computational burdens is splitting images into smaller, overlapping patches. However, this strategy conflicts with the detection of the other targets (e.g., the first-order images) which has a large aspect ratio.

\subsubsection{Large Aspect Ratio Challenge}

Non-zero order spectral traces exhibit an elongated morphology, with widths spanning $\sim$300–400 pixels\footnote{$\sim$300 for the first order spectral images. It could reach $\sim$700 pixels for higher orders.} and heights of $\sim$8 pixels, resulting in extreme aspect ratios ($\le$1:30). Splitting the original 9216$\times$9232 pixel image into smaller patches (e.g., 1024$\times$1024) risks truncating these extended features along the dispersion axis. This truncation introduces critical drawbacks: (1) During training, fragmented traces mislead the model, causing it to learn distorted or incomplete spectral features, which degrades detection accuracy. (2) During inference, reassembling truncated traces across patch boundaries becomes error-prone, as edge artifacts and misalignments corrupt the spectral continuity essential for parameter extraction. (3) Post-processing to reconcile fragmented objects demands additional computational steps (e.g., overlap stitching, duplicate removal), inflating pipeline complexity and runtime. To preserve the integrity of spectral traces and ensure reliable analysis, strategies must avoid truncation while balancing computational constraints inherent to gigapixel-scale slitless data.

\subsubsection{Information Asymmetry Challenge}
In slitless spectral images, the two directions (wavelength and spatial) don’t hold equal importance. Especially for stars, the wavelength direction is far more informative: it captures critical characteristics like absorption/emission line profiles and the broad continuum shape of the spectrum, which directly inform key parameters like temperature, surface gravity ($log \; g$), metallicity and other stellar properties. The spatial direction, on the other hand, mostly shows where the star is located and how the telescope blurs its light. While this helps pinpoint stars in crowded areas, it doesn’t reveal much about the star’s actual properties. This imbalance means analyzing slitless data requires focusing on the rich spectral details in the wavelength direction to unlock the science, even though both directions are part of the same image.

\subsubsection{Ambiguous Boundary Challenge}

A fundamental challenge that distinguishes astrophysical object detection from common computer vision tasks is the nature of object boundaries. In typical datasets (e.g., COCO, ImageNet), objects like cars, buildings, or people possess physically well-defined and sharp edges, allowing for unambiguous annotation of their bounding boxes. This high-contrast separation between the object and the background simplifies both the training and evaluation processes.

In contrast, spectral traces in astronomical images inherently lack such clear-cut boundaries. The light from a star or galaxy, dispersed by the grating and shaped by the telescope's PSF, fades gradually into the sky background. The flux diminishes smoothly, particularly in the faint outer wings of the trace, creating a "fuzzy" or probabilistic boundary rather than a distinct edge. This makes the task of defining a precise ground-truth bounding box intrinsically ambiguous. A human annotator or an automated algorithm must make a subjective decision about where the signal effectively ends and the background begins.

This ambiguity has direct consequences for the deep learning pipeline. During training, the model is exposed to ground-truth labels that contain this inherent uncertainty, which can affect the convergence of localization loss. More critically, it complicates the evaluation process. Metrics like Intersection over Union (IoU) are highly sensitive to the exact placement of bounding box edges. A predicted box might be scientifically accurate in encompassing the source's flux but could be penalized with a low IoU score simply because it differs slightly from the equally valid but subjective ground-truth annotation. This challenge requires careful consideration when interpreting performance metrics and highlights a key difference from standard object detection problems.

\subsubsection{The background challenge} 

Our primary objective is to detect zero- and first-order spectral images in slitless spectroscopy data. For bright stars, higher-order spectra (e.g., second, third, or negative first orders) are classified as background to avoid interference. For faint stars, however, all order spectral traces are designated as background due to two critical factors: (1) their low signal-to-noise ratio (SNR) renders spectral extraction unreliable even if detected, and (2) their inclusion would  increase overlapping regions between bounding boxes, introducing detection ambiguities that degrade model performance. This results in a highly complex background that shares morphological and spectral features with the target signals, posing unique challenges for distinguishing true sources from noise. The similarity between targets and background necessitates advanced methods to mitigate false positives while preserving detection sensitivity.

\vspace{10pt}

In summary, the primary challenges of the task stem from the unique characteristics of CSST slitless spectral data. A key difficulty is detecting targets at extreme scales within ultra-high-resolution images ($\sim$9000$\times$9000 pixels), where compact zero-order spectra are minuscule ($\sim$10$\times$10 pixels) while first-order spectra are highly elongated features with extreme aspect ratios ($>$30:1). This disparity creates an information asymmetry, as the feature-rich wavelength direction holds far more diagnostic power than the spatial direction. The task is further complicated by a complex background containing higher-order spectra and faint sources, which can mimic the targets and cause detection ambiguities. Unlike typical computer vision objects with sharp edges, these spectral traces have ambiguous, ``fuzzy'' boundaries that fade into the background, making precise annotation and evaluation difficult. Simultaneously, the framework also faces the challenge of spectral blending in crowded fields. This confluence of issues creates a paradoxical demand for a model that is both complex enough to disentangle these varied and overlapping signals, yet lightweight enough to handle the immense computational load of the gigapixel-scale images without training instability.

\subsection{the detection models}
Object detection frameworks are broadly categorized into two-stage and one-stage approaches. Two-stage models, such as the R-CNN family (e.g., \citealt{girshick14CVPR}, \citealt{girshickICCV15fastrcnn}, \citealt{renNIPS15fasterrcnn}, \citealt{he2017mask}), employ a hierarchical process: first, a Region Proposal Network (RPN) identifies potential object regions (bounding boxes), then a second network refines these proposals by classifying objects and adjusting coordinates. This two-step refinement enhances accuracy, but increases computational complexity and latency. In contrast, one-stage models like YOLO (You Only Look Once, \citealt{redmon2016you}), SSD (Single Shot Object MultiBox Detector, \citealt{liu2016ssd}), and RetinaNet (\citealt{lin2017focal}) streamline detection by directly predicting object classes and bounding boxes from a grid of predefined anchor boxes, eliminating the proposal step. This design prioritizes speed and efficiency, enabling real-time performance, though at the cost of reduced precision in cluttered scenes.

One-stage models excel in speed-critical applications (e.g., autonomous driving) but face challenges like class imbalance\footnote{It refers to class imbalance between targets and background. Advanced one-stage models could mitigate this through specialized techniques.} due to dense anchor sampling. Two-stage models, while slower, achieve higher accuracy by filtering irrelevant proposals early, making them ideal for precision-focused tasks (e.g., medical imaging). The choice hinges on balancing accuracy needs against computational constraints, with two-stage methods favoring detailed analysis and one-stage prioritizing rapid inference.

Initially, we prioritized two-stage object detection models, such as Mask R-CNN, due to their superior accuracy in bounding box localization and segmentation. This preference aligned with our task requirements, which prioritize precision over real-time processing. Mask R-CNN, an extension of Faster R-CNN, enhances detection by generating pixel-level segmentation masks alongside bounding boxes, a feature proven effective for star and galaxy detection in multi-band astronomical imaging \citep{Burke_2019}. However, training Mask R-CNN on our high-resolution slitless spectral data ($\sim$9000$\times$9000 pixels) exceeded our computational resources, as its two-stage architecture (region proposal followed by refinement) demands significant GPU memory and processing power.

We subsequently shifted focus to one-stage detectors, evaluating RetinaNet and YOLO. While RetinaNet’s focal loss mechanism addresses class imbalance—a critical issue in our data—its reliance on image tiling (splitting large images into smaller patches) introduced new challenges. Tiling fragmented extended spectral traces across patch boundaries, compromising target completeness and complicating reassembly. For slitless spectroscopy, where spectral continuity is essential for parameter extraction, fragmented detections degrade scientific utility.

YOLO, in contrast, processes full-resolution images without tiling, preserving the integrity of elongated spectral traces and minimizing edge artifacts. Its anchor-based architecture, optimized for small objects ($\sim$10$\times$10 pixel zero-order spectra), and efficient use of computational resources made it uniquely suited to our needs. While YOLO’s simplified loss function may trade some precision compared to two-stage methods, its ability to maintain target completeness and operate within hardware constraints rendered it the optimal choice for our high-resolution slitless spectral data.
Further more, although this work focuses on robust detection of spectral traces, our long-term goal is to develop an end-to-end framework capable of extracting astrophysical parameters (e.g., temperature, metallicity) directly from 2D slitless spectra. YOLO’s modular design and extendable head architecture also align with this objective.

A review of YOLO's history, structure, and architecture is provided in Appendix \ref{sec:YOLO}. This paper employs YOLOv5 \citep{JocherYOLOv5} and YOLO11 \citep{JocherYOLO11}. Briefly, YOLO11 offers increased accuracy and versatility, particularly for sub-class classification, but requires more computational resources and is more complex. YOLOv5, on the other hand, prioritizes speed and ease of use, demanding less computation, but does not offer sub-class classification. Therefore, we selected YOLOv5 for detecting numerous targets in dense areas, while YOLO11 was used for classifying objects into sub-classes (e.g., stars and galaxies) in moderately dense areas.

While classical source detection algorithms like SExtractor \citep{sex1996} are powerful for direct imaging, they are not well-suited to handle the fundamental challenges of slitless spectroscopic data. A primary difficulty is the detection and deblending of sources, as non-zero order spectral traces exhibit highly elongated morphologies with extreme aspect ratios and often overlap in crowded fields. Legacy pipelines, which are typically optimized for isolated sources, often exclude these overlapping spectral regions to simplify analysis, which leads to a sacrifice of critical astrophysical information. Furthermore, these traditional methods struggle to correctly associate the dispersed first-order spectra with their corresponding zero-order images without prior positional information from contemporaneous imaging. This issue of spatial uncertainty is a key challenge for the CSST survey, where the absence of direct source localization can prevent the precise extraction of spectra. Our YOLO-based framework is framed as a novel approach designed specifically to overcome these intrinsic limitations by learning the complex spatial and spectral features holistically and directly from the 2D slitless images.

\vspace{10pt}
This paper is organized as follows: Section \ref{sec:data} introduces the simulated CSST slitless spectroscopy dataset. Section \ref{sec:training_process_and_evaluation} details the data preprocessing pipeline, training protocols, and evaluation metrics used to assess model performance. Section \ref{sec:results} presents the detection results, including precision-recall metrics, robustness in crowded fields, and applications like quasar finding. Section \ref{sec:discussion} discusses bounding box offset and detection limits. Section \ref{sec:conclusion} summarizes key findings and outlines future work. Appendix \ref{sec:YOLO} provides details on the history, structure, and architecture of YOLO models, and Appendix \ref{sec:appendix_figure} contains supplementary figures.

\section{Data}
\label{sec:data}
Two datasets are included in this paper, for different purposes. The simulated CSST data used to train and validate our YOLO models are generated by a high-fidelity, comprehensive simulation suite designed to produce realistic, pixel-level mock observations. This ensures that our training data accurately represent the conditions of real CSST observations. The simulations incorporate detailed and validated models of the telescope's key characteristics. The Point Spread Function (PSF) is derived from a full optical system simulation that includes both static and dynamic aberrations, with its accuracy confirmed against laboratory measurements. For spectroscopy, the dispersion curves are modeled using the established aXe methodology \citep{axe2009}, with position-dependent polynomials that account for CSST's unique bidirectional dispersion and field rotation. Furthermore, a comprehensive suite of instrumental and observational effects is included, such as detailed models for stray light (including Earthshine and Zodiacal light), cosmic rays, and critical CCD effects like the brighter-fatter effect and charge transfer inefficiency (CTI), both of which were validated against established methods. The realism of these simulated datasets ensures that our deep learning model is trained on data that closely mimics the complex realities of the CSST survey, validating its robustness for scientific application. More details could be found in \cite{weichengliang2025} and \cite{zhangxin2025}.

\subsection{Simulation A: Different Densities, Stars Only}

Our first objective is to detect both zero-order and first-order slitless spectral traces in fields of varying stellar densities. Upon successful detection, only the positional coordinates of these traces are transmitted to the CSST data processing pipeline. The astrophysical origin of the spectra—whether from stars, galaxies, or other sources—is not prioritized at this stage, as the pipeline’s immediate requirement is accurate spatial localization rather than spectral classification. 

To achieve this objective, we employ Simulation A, a synthetic dataset designed to replicate slitless spectral observations across diverse Galactic environments. The simulation comprises stellar catalogs spanning distinct regions of the Milky Way, characterized by varying stellar densities. For instance, two representative fields include regions centered at Galactic longitude $l=90^{\circ}$ with latitudes $b=10^{\circ}$ and $b=20^{\circ}$, chosen to sample both the Galactic plane (high density) and higher Galactic latitudes (lower density).

Given the uncertain distribution of extragalactic sources (e.g., galaxies) in these regions, the simulation currently focuses exclusively on stellar populations, generated using the Galaxia \citep{Sharma_2011} synthetic Milky Way model . Each simulated field is labeled according to its Galactic coordinates (e.g., mwl90b10 denotes $l=90^{\circ}, b=10^{\circ}$), enabling systematic analysis of detection performance as a function of stellar crowding. This approach isolates the impact of source density on spectral trace detection, providing a controlled framework to optimize pipeline robustness for CSST’s anticipated observing scenarios.

This simulation leverages an older version of the CSST instrument model, incorporating all instrumental effects except the 16-channel bias and sky background.

\subsection{Simulation B: with Three Kinds of Sources}
Our second objective focuses on detecting and classifying astrophysical sources (stars, galaxies, and quasars) within slitless spectral images. To achieve this, we employ Simulation B, an updated synthetic dataset meticulously aligned with the CSST’s observational survey strategy. The observed region centered at around RA=170$^\circ$, Dec=-23.5$^\circ$. This simulation incorporates full instrumental effects, and some of them have been corrected, such as the 16-channel bias, flat-field correction, and dark current, by the CSST data processing pipeline. The simulation includes three types of sources: stars, galaxies, and quasars.

\vspace{10pt}

The instrument model in Simulation B differs from that of Simulation A, most notably in PSF, which defines how light from a point source spreads across the detector. In Simulation B, the updated PSF has a narrower spatial profile compared to Simulation A, reducing the footprint of light from individual sources. 
Taking CCD 01 with its two gratings as an illustration, Simulation A showed a PSF Full Width at Half Maximum (FWHM) of $\sim$4 pixels for both. In Simulation B, these values diverged to $\sim$4 pixels for one grating and $\sim$2 pixels for the other.

The PSF, instrumental effects, and other parameters within the CSST simulation are not designed to be freely configured by the user. Our research was conducted over a long period, during which the simulated PSF was updated to reflect hardware revisions. Consequently, Simulation A corresponds to an older version of the PSF, while the PSF used in Simulation B is based on the most recent results. The older version of the PSF spread wider than the new version. We retained the older version due to the need to test our algorithm's robustness. For the instrumental effects, their specific quantitative values cannot be altered; however, each effect can be individually enabled or disabled during the simulation process.

Our methodology utilizes two distinct simulation sets to comprehensively test our algorithm's performance and robustness. Simulation A provides a cleaner, idealized dataset by intentionally omitting the 16-channel bias and sky background effects, which can cause gradients across the detector mosaic. This simpler configuration serves as a controlled environment to validate the core functionality of our detection algorithm. In contrast, Simulation B is designed to be more similar to a real observation by including a full suite of instrumental effects and applying partial corrections. The decision to use both configurations was driven by a need for diversity in our testing data and was also a pragmatic choice, as the official correction pipelines were unavailable when the project began. Subsequent analysis confirmed this approach was valid, demonstrating that the final detection metrics were not significantly affected by the inclusion or exclusion of these specific, well-understood instrumental effects.

\section{Training Process and Evaluation}
\label{sec:training_process_and_evaluation}

\subsection{Data Processing}

Data processing plays a critical role in optimizing model performance, beginning with contrast enhancement to improve YOLO’s feature extraction capabilities. To mitigate image elongation and reduce computational demands, we divide images into 10 vertical strips, rescale them to square dimensions, and thereby decrease the number of detection targets per image by 10\%, lowering resource consumption per training epoch. These preprocessing steps collectively refine input quality, accelerate training convergence, and enhance detection accuracy.

\subsubsection{Enhance
Image Contrast}

Although the average pixel value in Simulation A's original image hovers around 5, cosmic rays can induce extreme, localized increases, pushing some values to 30,000. These significant discrepancies necessitate thorough data cleaning and calibration before analysis. Crucially, normalizing these outliers prevents them from skewing the model's results, leading to a more reliable and accurate detection system.

To improve image normalization and contrast, we tested multiple techniques such as ZScale, Histogram Equalization, and Contrast Limited Adaptive Histogram Equalization (CLAHE). After rigorous analysis, we adopted ZScale—a method rooted in astronomical imaging and based on the IRAF zscale algorithm. ZScale optimizes contrast by dynamically calculating the most suitable minimum and maximum intensity values, which sharpens critical features and boosts overall image clarity. This approach not only refines visual quality but also strengthens the precision of our detection framework. Based on our experimental results, we have determined that image nomarlization is a critical prerequisite for the model to achieve convergence during training.

\begin{figure}[h]
\centering
\includegraphics[width=\columnwidth]{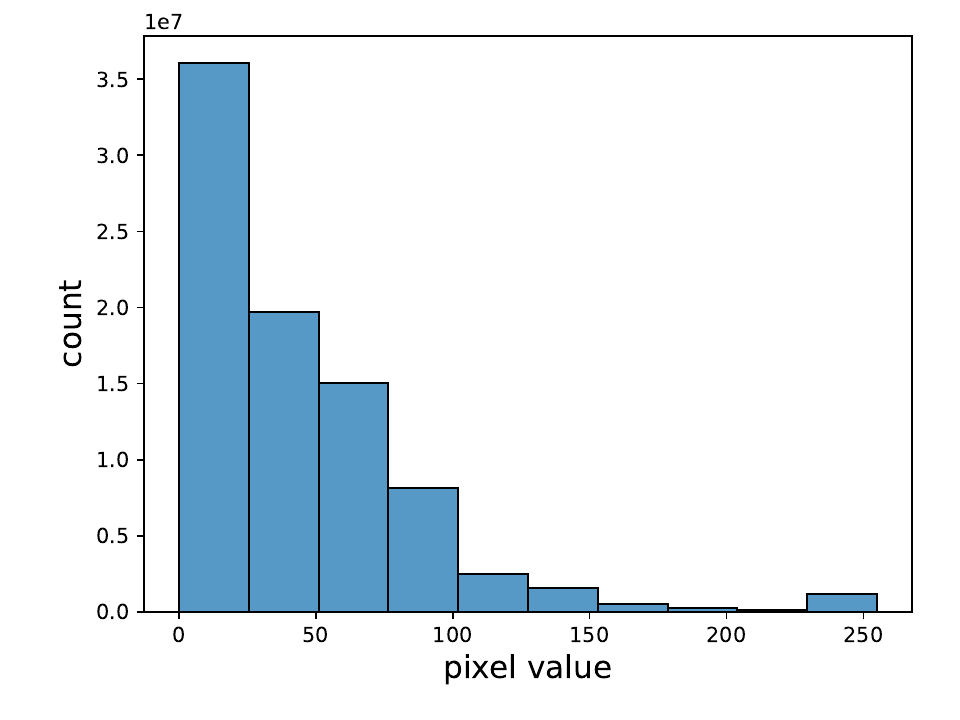}
\centering\caption{This figure illustrates the pixel intensity distribution of a ZScale-normalized slitless spectral image from Simulation A mwl90b20, Chip 01 (GI). The ZScale algorithm, a standard tool in astronomical image processing, optimizes contrast by dynamically adjusting pixel values to highlight critical spectral features while suppressing noise. By stretching the intensity range to emphasize subtle variations, this method enhances the visibility of faint or overlapping spectral traces, enabling clearer identification and analysis of the data’s structural details.}
\label{fig:pixel_intensity}
\end{figure}

Following ZScale processing, pixel intensities were rescaled from the range [0, 1] to [0, 255] by multiplying values by 255. The resulting distribution of pixel intensities across the dataset, shown in Figure \ref{fig:pixel_intensity}, indicates a significant prevalence of low-intensity pixels. The histogram also displays a rapid decline in pixel count with increasing intensity, although a tail towards higher values suggests the presence of some brighter elements within the images.

\subsubsection{Image Size Rescale}

\begin{figure}[h]
\centering
\includegraphics[width=\columnwidth]{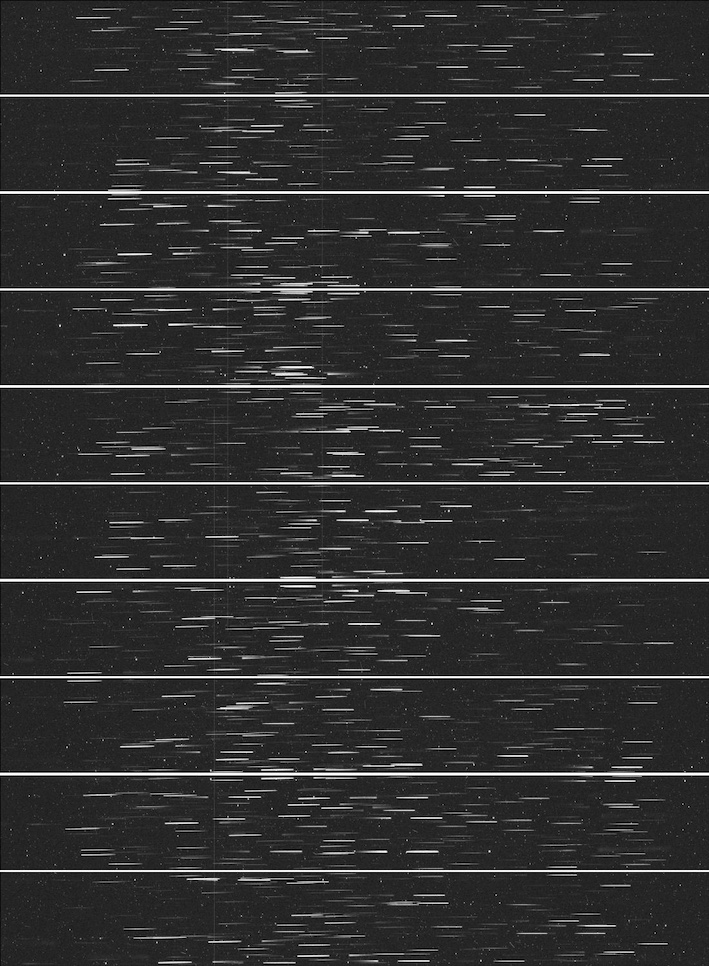}
\centering\caption{This figure illustrates the process of splitting the original image into strips. The original image size is 9216$\times$9232 pixels. We divided the image into 10 sub-images, each with a size of 9216$\times$979 pixels. To ensure continuity and avoid losing any information at the edges, we included an overlapping area of 9216$\times$62 pixels between each sub-image. This overlapping region helps maintain the integrity of the data and ensures that all relevant features are captured in the sub-images. By splitting the image in this manner, we can manage the large image size more effectively and facilitate more efficient processing and analysis.}
\label{fig:train_img_0}
\end{figure}

The PSF has a FWHM of $\sim$4 pixels. In the original 9216$\times$9232 images, class A objects appear elongated, while class B objects are compact. Their bounding are determined by the spectral traces predefined in the simulation’s configuration file, with a vertical and horizontal adjustment of $\sim \pm$20 pixels to ensure complete target coverage.
To address this and optimize model performance, we implement a horizontal cropping strategy. As vertical cropping would disrupt class A’s elongated structure, we instead split the image horizontally into 10 sub-images, each sized 9216$\times$979 pixels. To avoid truncating targets, adjacent sub-images share a 62-pixel overlapping region, guaranteeing every unabridged object remains fully captured. These cropped sub-images (visualized in Figure \ref{fig:train_img_0}) are then rescaled to standardize their dimensions for model training.

YOLOv5 requires input images to be square and divisible by the model's stride values. For YOLOv5-P5 (3 output layers: P3, P8, P16 at strides 8, 16, 32) and YOLOv5-P6 (4 output layers: P3, P4, P5, P6 at strides 8, 16, 32, 64), we prioritized high-resolution inputs to reduce information loss. However, balancing computational efficiency, we selected a 4544×4544 pixel size as the optimal compromise.

\begin{figure*}[t]
\centering
\includegraphics[width=2\columnwidth]{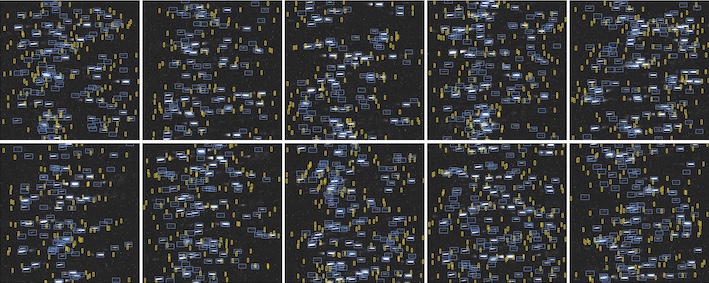}
\centering\caption{The 9216$\times$979 pixel strips (Figure \ref{fig:train_img_0}) were resized to a square 4544$\times$4544 format, enhancing adaptability to processing pipelines while preserving the structural integrity of key features. This resizing strategy achieves an optimal trade-off: retaining sufficient resolution for precise target detection while minimizing computational overhead. Zero-order (B, yellow) and first-order (A, blue) spectral traces are annotated to aid classification, streamlining analysis within the standardized sub-images. Stars fainter than magnitude 22 are considered as background noise, ensuring focused detection of relevant targets. Each resized sub-image contains approximately 220 objects.}

\label{fig:train_img_1}
\end{figure*}

The original 9216×979 sub-images (generated via horizontal cropping, as described earlier) were resized to 4544×4544 pixels. These rescaled images (Figure \ref{fig:train_img_1}) preserve critical features while reducing computational load. Each sub-image contains approximately 220 objects (Class A and Class B). 

This approach ensures compatibility with YOLO’s architecture while maintaining detection accuracy for elongated (Class A) and compact (Class B) targets.

The sizes of bounding boxes vary depending on the source direct imaging position and spectral traces. After rescaling, for Class A objects, the width ranges from approximately 230 to 260 pixels, while the height differs based on their location: objects with direct imaging position in the left 2/5 have a height of $\sim$146 pixels, whereas those in the right 3/5 are $\sim$130 pixels tall. This disparity arises because the left and right spectral traces originate from separate gratings, producing distinct scattering morphologies. Class B objects (blue) exhibit consistent dimensions, with a height of $\sim$130 pixels and a narrower width of $\sim$40 pixels. These variations highlight the influence of grating geometry and spectral scattering patterns on object morphology.

\subsubsection{Bounding Box Annotation}

The process for generating ground-truth annotations begins with a baseline calculation from the aXe software, which uses a CSST-specific configuration file to determine the starting ($x_1, y_1$) and ending ($x_2, y_2$) pixel coordinates of a spectral trace for an idealized point source. From these two endpoints, a preliminary bounding box is defined by taking the minimum and maximum of these coordinates, such that $x_{min}=min(x_1,x_2)$, $x_{max}=max(x_1,x_2)$, $y_{min}=min(y_1,y_2)$, and $y_{max}=max(y_1,y_2)$. However, this theoretical footprint is insufficient for realistic annotation for two primary reasons. First, as an initial geometric projection, the aXe output does not account for the Point Spread Function (PSF), which causes the light from even a point source like a star to spread across multiple pixels. Second, our targets include not only point sources but also extended sources, such as galaxies, which have intrinsically larger spatial profiles. The combination of the PSF's effect and the inherent size of extended sources results in a spectral trace with an ambiguous or "fuzzy" boundary, where the object's flux diminishes gradually into the sky background. To ensure the annotation robustly captures the entire dispersed signal and compensates for these effects, this initial bounding box is expanded outward by a fixed margin of 14 pixels on all four sides, creating the final region used for training and evaluation.

\subsubsection{Noise or Signal: Brightness Thresholding}
Our objective is to detect zero-order (Class B) and first-order (Class A) spectral traces without prior knowledge of source positions. However, observational data contains numerous faint sources (evident in the magnitude distribution). Including all such sources would result in excessively dense bounding boxes, overwhelming the model during training and complicating performance evaluation. To address this, we only label zero- and first-order traces from bright sources as ground truth. All other spectral orders from bright sources, along with all traces from faint sources, are treated as background.

Bright and faint sources are distinguished using a magnitude threshold. The magnitude is calculated by integrating flux within the designated spectral bands: GU (255–420 nm), GV (400–650 nm), and GI (620–1000 nm), without applying filter corrections. To balance source density and minimize overlapping targets, we empirically set the threshold at m=22 for the GI band in Simulation A, mwl90b20. This value was then retained for consistency across other bands (GU, GV), though it may not represent the theoretically optimal cutoff.

\subsection{Train}
\subsubsection{Computational Resources}
Our server leverages 8 NVIDIA A100-PCIE-40GB GPUs, each equipped with 40 GB of HBM2e memory and a memory bandwidth of 1.6 TB/s to accelerate large-scale data processing. Built on NVIDIA’s Ampere architecture (compute capability 8.0), each GPU features 6,912 CUDA cores and 432 Tensor Cores for accelerating mixed-precision workloads, with base and boost clock speeds of 765 MHz and 1,410 MHz, respectively. Designed for efficiency, each GPU operates at a 250W thermal design power (TDP), balancing high-performance computation (e.g., deep learning training, astronomical dataset analysis) with energy demands.  

\subsubsection{Trainig Strategy}

To achieve an effective balance of efficiency and robustness in our model's training, we began by initializing our models with pre-trained weights from the Ultralytics YOLOv5 repository. Given that the unique morphologies of our astronomical targets—including compact zero-order images, elongated first-order spectra, and especially their indistinct boundaries—differ significantly from standard objects, a full retraining approach was implemented. By training all layers from these initial weights without freezing any, we enabled the network to fully adapt to our domain-specific data characteristics. The model's generalization capability was further enhanced through the use of default augmentation techniques.

YOLO's default augmentation pipeline—including mosaic synthesis, geometric transformations (rotation, flipping, scaling), and copy-paste operations—is retained to enhance model generalization despite generating synthetic spectral features unobserved in slitless spectroscopy. Although rotated or scaled spectra deviate from physical reality (where slitless dispersion fixes spectral scaling via wavelength dependence), these augmentations enforce invariance to observational variabilities such as pointing inaccuracies, background noise fluctuations, and spatial overlaps in crowded fields. By simulating diverse conditions, the model learns robust feature representations, reducing overfitting to fixed morphological artifacts. While introducing minor synthetic inconsistencies, retaining YOLO’s native augmentation framework ensures compatibility with its optimization protocols, achieving empirically validated performance gains without compromising spectral fidelity in downstream analysis.

\subsubsection{Dataset}

In Simulation A mwl90b20 for Chip 01 (GI), we analyzed 30 images generated from 30 observational pointings. Surprisingly, experiments revealed that training with only 1–3 images achieved robust detection accuracy while drastically reducing computational costs. This minimal training data approach highlights the effectiveness of our preprocessing pipeline—high-quality normalization, cropping, and rescaling enable the model to generalize well even with limited input. The results demonstrate that optimized workflows, rather than sheer data volume, drive performance, making our method both resource-efficient and scalable for large-scale astronomical datasets.

Therefore we allocated three original images for training, one for validation, and 26 for testing. Each image was cropped into 10 sub-images during preprocessing, producing 30 training sub-images, 10 validation sub-images, and 260 testing sub-images.

\subsubsection{Models}

YOLOv5 provides two architectures, P5 and P6, each with five model sizes (Nano, Small, Medium, Large, XLarge) to balance speed and accuracy. The P5 variants (e.g., YOLOv5n) prioritize lightweight performance for real-time tasks, while P6 models (e.g., YOLOv5n6) enhance resolution handling for complex scenarios. In contrast, YOLO11’s architectural complexity imposes higher computational demands, particularly when processing large images ($\sim$4200$\times$4200 pixels) to retain target fidelity. Due to these constraints, only YOLO11n (Nano) and YOLO11s (Small) are operable on our server, avoiding CUDA memory limits while maintaining efficient processing and data integrity.

\subsection{Evaluation}

Intersection over Union (IoU) is a critical evaluation metric in object detection, measuring the overlap accuracy between a predicted bounding box and its corresponding ground truth. Calculated as the ratio of the overlapping area (intersection) to the combined area (union) of the two boxes, IoU values range from 0 (no overlap) to 1 (perfect alignment). Higher IoU scores reflect greater detection precision. To rigorously assess model performance, an IoU threshold is set to classify detections: the standard threshold of 0.5 is often relaxed, but in our work, we enforce a stricter threshold of 0.7 (it reduces false positives but may increase false negatives). This ensures higher confidence in detections, and prioritizing precise localization—a necessity for applications requiring fine-grained accuracy.

Evaluating an object detection model involves several key metrics and techniques to ensure its accuracy and effectiveness. We use 4 key parameters to evaluate models, which are precision, recall, AP (Average Precision) and F1-score. 

Precision and recall are crucial metrics for evaluating the performance of a model, especially in the context of classification tasks.
Precision measures how accurate the positive predictions are. It is calculated as the ratio of true positive predictions to the total number of positive predictions (true positives + false positives). In other words, precision answers the question: “Of all the instances that the model predicted as positive, how many were actually positive?” High precision indicates that the model makes few false positive errors. Recall, on the other hand, measures the model’s ability to identify all relevant instances within a dataset. It is the fraction of relevant instances that were retrieved. It is calculated as the ratio of true positive predictions to the total number of actual positives (true positives + false negatives). Recall answers the question: “Of all the actual positive instances, how many did the model correctly identify?” High recall indicates that the model makes few false negative errors.

Understanding the trade-off between precision and recall is essential. A model with high precision but low recall is very selective in its positive predictions, which means it misses many actual positives (high false negatives). Conversely, a model with high recall but low precision captures most of the actual positives but also includes many false positives.

Average Precision (AP) is a crucial metric for assessing the performance of object detection models. It provides a comprehensive evaluation of a model’s ability to accurately detect and localize objects within images. AP is calculated as the area under the precision-recall curve for each class, which is then averaged across all classes. This consolidated score reflects the model’s overall detection accuracy and effectiveness. 
The significance of AP lies in its ability to offer a single, unified measure that captures both precision and recall across different thresholds. This makes it an invaluable tool for comparing the performance of various object detection algorithms. AP is extensively used in benchmark tests and competitions, serving as a standard metric for evaluating and ranking models. Its widespread adoption underscores its importance in the field of computer vision, where accurate object detection is critical for applications ranging from autonomous driving to medical imaging. 

The F1-Score is a crucial metric that represents the harmonic mean of precision and recall. It provides a balanced evaluation of a model’s performance by considering both the accuracy of positive predictions (precision) and the ability to identify all relevant instances (recall). This metric is particularly useful in scenarios where there is an uneven class distribution or when the cost of false positives and false negatives is high. By combining precision and recall into a single score, the F1-Score offers a comprehensive view of the model’s effectiveness, ensuring that neither metric is overlooked. This makes it an invaluable tool for assessing the robustness and reliability of classification models.

\section{Results}
\label{sec:results}
\subsection{YOLOv5, Simulation A, star only images}
\subsubsection{Chip 01 (GI)}
In this section, we present a detailed analysis of the performance of YOLOv5 models of varying sizes. Table \ref{tab:model_structure} provides a comprehensive overview of the number of layers, parameters, and the computational complexity of each model. The computational complexity is measured in GFLOPs (Giga Floating Point Operations per Second), which quantifies the number of billion operations the model can perform per second. This metric is crucial for understanding the computational demands of each model variant.

Additionally, Table \ref{tab:model_performance} showcases the performance metrics of these models, allowing for a clear comparison of their efficiency and effectiveness in different scenarios. This thorough evaluation aims to guide users in selecting the most suitable YOLOv5 model based on their specific requirements and constraints.

\begin{table*}[t]
  \begin{center}
    \caption{number of layers and parameters}
    \label{tab:model_structure}

    \begin{tabular}{c|c|c|c|c|c} 
    \hline
      model & v5n & v5s & v5m & v5l & v5x \\
      \hline

      layers & 157 & 157 & 212 & 267 & 322 \\
      parameters & 1,761,871 & 7,015,519 & 20,856,975 & 46,113,663 & 86,180,143 \\
      complexity & 4.1 & 15.8 & 47.9 & 107.7 & 203.8 \\
        \hline
      model & v5n6 & v5s6 & v5m6 & v5l6 & v5x6 \\
      \hline
      layers & 206 & 206 & 276 & 346 & 416 \\
      parameters & 3,089,188 & 12,312,052 & 35,254,692 & 76,126,356 & 139,980,484\\
      complexity & 4.2 & 16.1 & 48.9 & 109.9 & 207.9 \\

    \end{tabular}
 
\vspace{10pt}
\footnotesize{
    \textbf{Note}: Complexity is measured in GFLOPs (Giga Floating Point Operations), representing the computational workload for one inference pass. P6 models are optimized for higher-resolution inputs, adding layers and parameters compared to P5 counterparts. Larger models (v5x/v5x6) prioritize accuracy at the cost of computational resources, while smaller variants (v5n/v5n6) favor speed and efficiency.
}
\end{center}

\end{table*}

\begin{table*}[t]
  \begin{center}
    \caption{Performance Evaluation of YOLOv5 Models on Simulation A mwl90b20, Chip 01 (GI) – Beams A and B}
    \label{tab:model_performance}
    \begin{tabular}{c|cccc|cccc} 
      \hline
      \multirow{2}{*}{\textbf{Model}} & \multicolumn{4}{c|}{\textbf{Beam A}} & \multicolumn{4}{c}{\textbf{Beam B}} \\
      & $\mathbf{F1_{max}}$ & $\mathbf{Recall}$ & $\mathbf{Precision}$ & $\mathbf{AP}$ & $\mathbf{F1_{max}}$ & $\mathbf{Recall}$ & $\mathbf{Precision}$ & $\mathbf{AP}$ \\
      \hline
      v5n & 0.937 & 0.909 & 0.966 & 0.959 & 0.917 & 0.888 & 0.947 & 0.950 \\
      v5s & 0.961 & 0.950 & 0.973 & 0.975 & 0.918 & 0.881 & 0.959 & 0.956 \\
      v5m & 0.956 & 0.945 & 0.969 & 0.970 & 0.926 & 0.895 & 0.959 & 0.955 \\
      v5l & 0.956 & 0.936 & 0.976 & 0.971 & 0.893 & 0.854 & 0.936 & 0.947 \\
      v5x & 0.960 & 0.950 & 0.969 & 0.972 & 0.928 & 0.899 & 0.959 & 0.962 \\
      \hline
      v5n6 & 0.941 & 0.932 & 0.950 & 0.964 & 0.920 & 0.891 & 0.951 & 0.955 \\
      v5s6 & 0.956 & 0.943 & 0.970 & 0.970 & 0.940 & 0.918 & 0.963 & 0.967 \\
      v5m6 & 0.909 & 0.920 & 0.898 & 0.947 & 0.910 & 0.861 & 0.965 & 0.949 \\
      v5l6 & 0.917 & 0.946 & 0.890 & 0.960 & 0.901 & 0.840 & 0.972 & 0.953 \\
      v5x6 & 0.949 & 0.946 & 0.951 & 0.973 & 0.914 & 0.873 & 0.960 & 0.954 \\
      \hline
      \textbf{Mean} & 0.944 & 0.938 & 0.951 & 0.966 & 0.917 & 0.880 & 0.957 & 0.955 \\
      \hline
    \end{tabular}
  \end{center}
  
  \vspace{10pt}
  \footnotesize{
    \textbf{Note}: Metrics computed at IoU threshold 0.7. $F1_{max}$ denotes the peak F1-score (harmonic mean of precision and recall) across all confidence thresholds, with adjacent columns displaying the corresponding precision and recall, while AP is derived as the area under the precision-recall curve. 
    
    The negligible performance gap between compact and large YOLOv5 models in this detection task starkly highlights that model size is not the bottleneck. Smaller models (e.g., YOLOv5n/v5s) achieve near-ceiling accuracy because homogeneous, high-contrast targets require limited feature complexity, while larger models (e.g., YOLOv5x) face diminishing returns due to over-parameterization and limited training data. This suggests computational resources are better invested in optimizing preprocessing or expanding dataset diversity rather than scaling model size. However, larger models may still prove valuable if future tasks involve fainter targets, overlapping spectra, or multi-class detection requiring nuanced feature extraction.} 
    
\end{table*}

\begin{figure*}[t]%
    \centering
    \begin{subfigure}[\centering Chip 01 (GI), ground truth]{{\includegraphics[width=0.8\columnwidth]{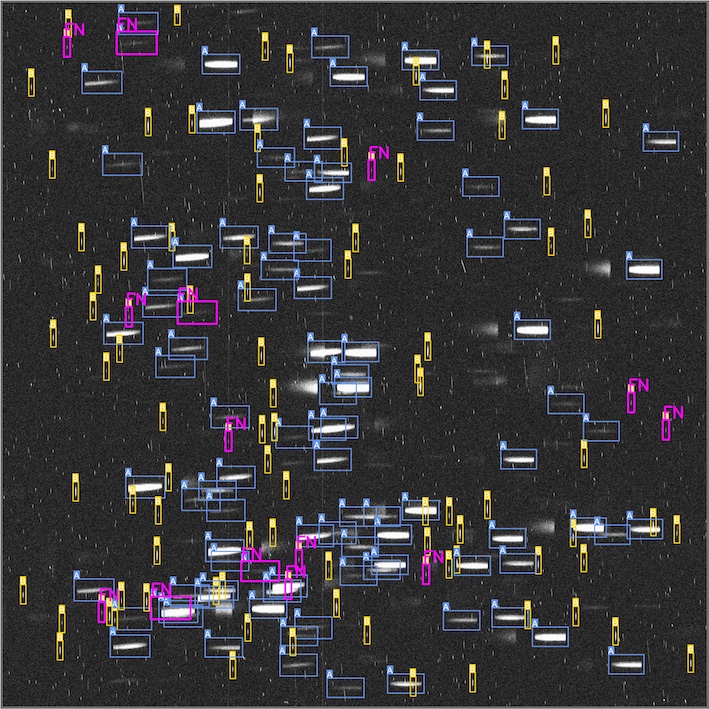}}}%
    \label{}
    \end{subfigure}
    \qquad
    \begin{subfigure}[\centering Chip 01 (GI), prediction]{{\includegraphics[width=0.8\columnwidth]{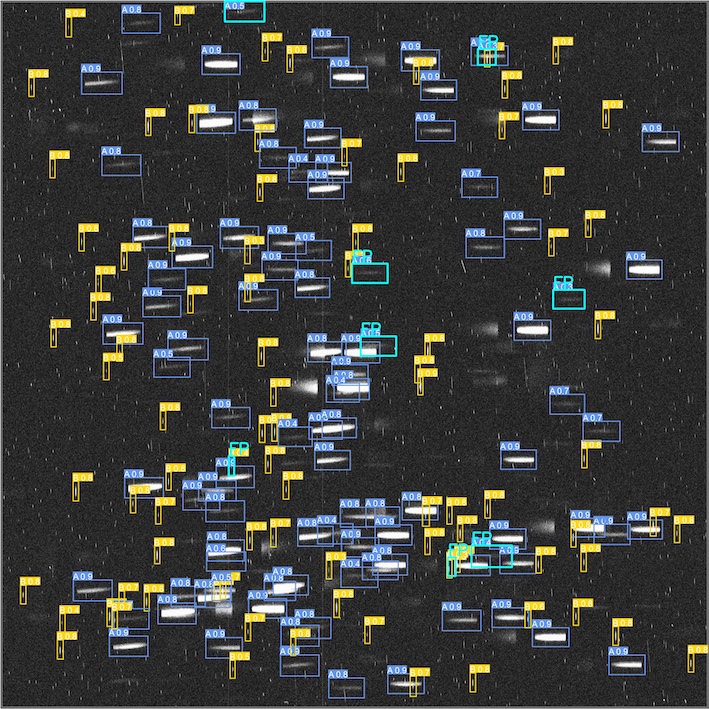}}}%
    \label{}
    \end{subfigure}
    \caption{(a) Ground truth (with a magnitude threshold of 22, targets fainter than the threshold are labeled as background) and (b) predictions from model YOLOv5s (with scores $\ge$ 0.25). Class A, marked by the blue bounding box, represents the first-order slitless spectral image. Class B, marked by the yellow bounding box, corresponds to the zero-order slitless spectral image. The differences between (a) and (b) are highlighted by magenta (false negatives) and cyan (false positives) boxes. Class A is our primary target. When achieving the maximum F1-score of 0.961, the recall and precision values are 0.950 and 0.973, respectively. This indicates that the model is highly effective in identifying true positives while maintaining a low rate of false positives and false negatives. The high precision value of 0.973 reflects the model’s ability to accurately identify objects within Class A, while the recall value of 0.950 demonstrates its proficiency in capturing the majority of relevant instances.}
    
    \label{fig:ground_truth_and_pred}
\end{figure*}

Our research indicates that the performance of various YOLOv5 model sizes is remarkably similar. Interestingly, in certain scenarios, simpler models outperform their more complex counterparts. This observation suggests that the increased complexity of larger models does not always translate to better performance. It would have several reasons.

Firstly, the nature of our dataset may inherently favor a simpler model. Given that the objects in our images are relatively easy to detect and distinguish, a more complex model may not be necessary. Secondly, models with more layers and parameters are prone to overfitting the training data, which can hinder their ability to generalize to new data. In contrast, a simpler model may offer better generalization. Thirdly, simpler models are easier to tune. Since the performance of a simple model (such as v5s) is already satisfactory, we do not need to invest significant effort in tuning a more complex model, which would also require a larger training set. Thus, in this specific task prioritizing lightweight models optimizes both training efficiency and deployment reliability. 

In our experiments, we utilized the YOLOv5 model to detect two distinct classes: A and B. We show the ground truth and the predictions of model YOLOv5s in Figure \ref{fig:ground_truth_and_pred}. The model demonstrated impressive performance, achieving a mean AP of 0.966 for all two classes at an IoU threshold of 0.7. For Class A, at the point where the F1-score reaches its maximum, the model achieved an impressive precision of 0.973 and a recall of 0.950. 

These results indicate that the YOLOv5 model is highly effective in accurately identifying and localizing objects within these two classes. 

While simpler models suffice for current tasks, more complex architectures may become essential for advanced challenges—such as detecting fainter targets, disentangling overlapping spectra, or handling multi-class detection and regression tasks—where nuanced feature extraction is critical.

\subsubsection{Chip 02 (GV) and Chip 03 (GU)}

In addition to training models for Chip 01, we also developed models for two other chips, Chip 02 (GV) and Chip 03 (GU). The data processing methodology for these chips follows the same procedures as previously described. However, for Chip 03, the sizes of the training, validation, and test sets differ due to the lower transmittance of the GU band, resulting in a lower sample density compared to the other two chips. Specifically, the sample sizes for training, validation, and testing are 23, 4, and 8, respectively.  After cropping each figure into 10 sub-figures, these sample sizes increase to 230 for training, 40 for validation, and 80 for testing. This adjustment ensures that the model has sufficient data to learn effectively, despite the lower initial sample density in the GU band. 

The left column presents the ground truth, while the right column features the predictions from the YOLOv5s model. These images originate from simulations of Chip 02 (GV) or Chip 03 (GU). In the ground truth figures, only targets with magnitudes $\le$22 in corresponding band are labeled.

For Chip 02 (GV), the balance of recall and precision for Class A are 0.906 and 0.964, respectively. For Class B, the corresponding values are 0.878 and 0.948. This indicates that the model performs well in detecting and identifying objects within these classes, maintaining a high level of accuracy and reliability. The first row of Fig.\ref{fig:ground_truth_and_pred_GVGU} shows an example of the comparison of the ground truth and predictions. 

For Chip 03 (GU), the balance of recall and precision for Class A are 0.965 and 0.913, respectively. For Class B, the corresponding values are 0.700 and 0.940. 

These results confirm that the YOLOv5s model reliably detects and classifies spectral traces across multiple bands, demonstrates promising potential for accurately locating spectral traces even without prior source location information.

\subsection{FP and FN}

A false positive refers to an error where the model erroneously flags a non-target object (e.g., noise or artifacts) as a valid detection. Conversely, a false negative occurs when the model overlooks a genuine target (e.g., a faint star or spectral trace) that exists in the data. Minimizing both errors is critical to enhancing the model’s precision (reducing false alarms) and recall (capturing true targets), ensuring robust and trustworthy performance of the detection model.

To evaluate detection errors, we compared model predictions (confidence score $\ge$ 0.25) against ground-truth annotations (with a magnitude threshold of 22). This identified false positive and false negative. Figure \ref{fig:fnfp}(a) visualizes these errors (false positives in cyan, false negatives in magenta) for the YOLOv5s model on Chip 01 spectral images of Simulation A mwl90b20. Figure \ref{fig:fnfp}(b) displays ground-truth objects with magnitude $\le$ 22, while Figure \ref{fig:fnfp}(c) expands the threshold to magnitude $\le$ 23, incorporating fainter, lower-magnitude objects. 

As shown in Figure \ref{fig:fnfp}a, false negatives are concentrated among faint objects with low magnitudes. The root cause of these missed detections lies in their diminished signal strength, which often falls below the model’s detection threshold due to their low brightness. Faint targets generate insufficient contrast against background noise or instrumental artifacts, causing the model to struggle to distinguish them from spurious signals. This limitation highlights the challenge of detecting low-magnitude objects in slitless spectroscopy, where faint sources are inherently harder to resolve. Future work could explore adaptive preprocessing or confidence score calibration to improve sensitivity to these marginal cases.

The analysis reveals two predominant sources of false positives (FPs) in our detection framework. First, faint objects excluded from ground-truth labels by the magnitude-22 threshold (Figure \ref{fig:fnfp}b) are misclassified as FPs when detected by the model. These faint detections (cyan in Figure \ref{fig:fnfp}a) become valid targets when the threshold is relaxed to magnitude 23 (Figure \ref{fig:fnfp}c), highlighting their dependence on annotation criteria. Second, edge-clipped objects—partially truncated by sub-image boundaries—are flagged as FPs during validation. Though excluded from training to avoid ambiguous labels, the model detects these partial objects, reflecting its sensitivity to incomplete features.

To balance evaluation practicality with scientific priorities, we retain the magnitude-22 threshold to prioritize high-confidence targets (e.g., bright stars, galaxies). While this approach inevitably penalizes the model for detecting faint sources, the trade-off ensures streamlined evaluation and alignment with observational goals, where precision for critical targets outweighs theoretical completeness. Introducing complex metrics (e.g., tiered thresholds) would complicate analysis without enhancing core objectives. Thus, we accept this limitation, emphasizing simplicity and focus over exhaustive detection of marginal or ambiguous cases.

In summary, false negatives arise when faint objects are too dim to detect; false positives occur when the model identifies valid faint sources excluded from ground-truth labels. This trade-off prevents overcrowded annotations but penalizes accurate detection of faint objects.

\subsubsection{In Dense Field Simulation A mwl90b10}
\begin{figure*}[t]%
    \centering
    \begin{subfigure}[\centering Chip 01 (GI), mwl90b10, ground truth]{{\includegraphics[width=0.8\columnwidth]{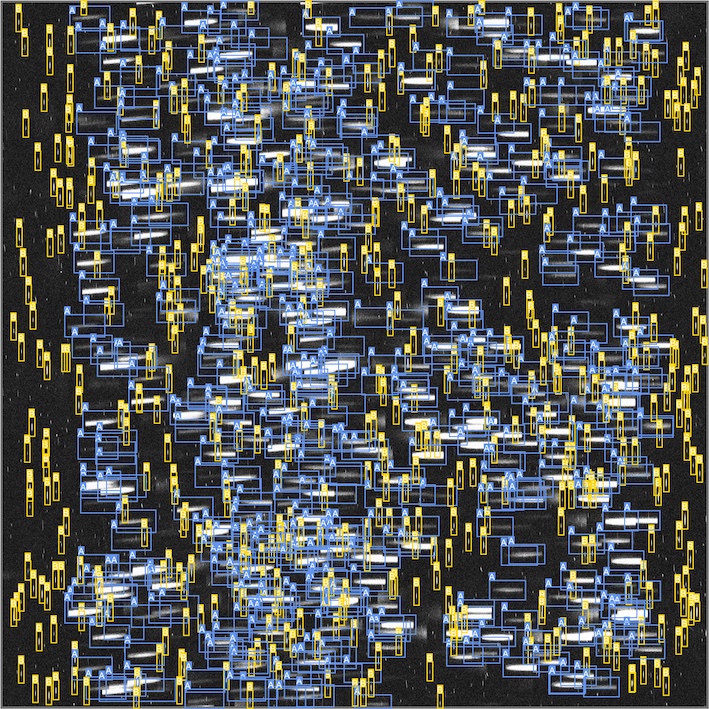} }}%
    \label{fig:YOLOv5s_b10_a}
    \end{subfigure}
    \qquad
    \begin{subfigure}[\centering Chip 01 (GI), mwl90b10, YOLOv5s predictions]{{\includegraphics[width=0.8\columnwidth]{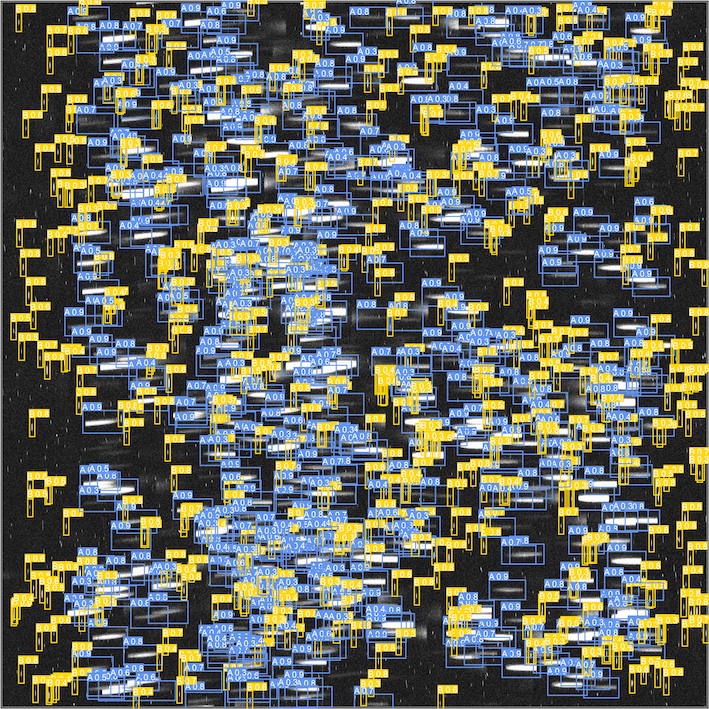}}}%
    \label{fig:YOLOv5s_b10_b}
    \end{subfigure}

    \caption{This figure demonstrates YOLOv5’s capability to detect spectral traces in densely populated star fields, using the Simulation A mwl90b10 as an example. Panel (a) illustrates the ground-truth annotations filtered by a magnitude threshold of 22. Panel (b) displays predictions from the YOLOv5s model trained on the mwl90b20 dataset. Despite significant overlap between bounding boxes—which complicates precise pairing of true and predicted detections for conventional metric evaluation (e.g., precision, recall)—visual inspection reveals that the model accurately identifies the zero-order (compact) and first-order (elongated) spectral images for nearly all bright sources. This qualitative assessment underscores YOLOv5’s robustness in dense stellar environments, confirming its suitability for astronomical tasks requiring reliable detection in crowded fields.}
    
    \label{fig:YOLOv5s_b10}
\end{figure*}

While high-density regions (e.g., Simulation A mwl90b10) inherently challenge precise detection due to overlapping slitless spectra, our results demonstrate the YOLOv5 model’s surprising effectiveness. As shown in Figure \ref{fig:YOLOv5s_b10}, the model trained solely on medium-density data (Simulation A mwl90b20) successfully identifies both zero-order and first-order spectral images in crowded fields, defying expectations. 

Though conventional metrics (e.g., precision, recall) are impractical here due to bounding-box overlaps, the visual clarity of detections underscores the model’s adaptability to complex scenarios.

This performance highlights YOLOv5’s robustness: despite being trained on a medium-density region, it generalizes to dense regions by prioritizing high-confidence targets. The figure reveals accurate localization of bright sources (magnitude $\le$22), even amid spectral overlaps that typically confound detection systems. While quantitative evaluation remains limited by annotation challenges, the visually compelling results validate the model’s potential for real-world applications in crowded stellar fields, where traditional methods falter.

\subsection{YOLO11: Extending Detection to Stars and Galaxies}

In prior subsections, we focused on identifying zero- and first-order slitless spectral traces of stellar sources under controlled conditions, analyzing the impact of source density (e.g., sparse vs. crowded fields in mwl90b20 and mwl90b10), detector characteristics (GI, GV, GU bands), and signal-to-noise ratios (via magnitude thresholds). To isolate foundational performance metrics, galaxies were excluded due to the lack of corresponding catalogs in the selected regions. This approach allowed us to systematically evaluate the YOLO framework’s capability to resolve spectral features in varying stellar environments without introducing complexities like extended galaxy morphologies or blended spectra.

Building on this foundation, we now expand the framework to address more practical observational scenarios by incorporating galaxies as distinct classes. This extension introduces four detection categories: (1)zero-order stars (compact, high signal-to-noise), (2)first-order stars (dispersed, wavelength-dependent),
(3)zero-order galaxies (slightly extended, resolved cores), and
(4)first-order galaxies (broad, low-surface-brightness spectra). Quasars are also included as the fifth category.

We initially adopted YOLOv5 for its computational efficiency and proven capability in processing high-resolution, target-rich slitless spectroscopy data—a critical requirement for our foundational pipeline. However, when expanding detection to finer astrophysical subclasses (e.g., distinguishing stars from galaxies), its architectural limitations became apparent: YOLOv5 struggles to differentiate spatially compact stellar spectra (smooth, point-like profiles) from extended galaxy spectra (irregular or structured profiles) due to insufficient mechanisms for resolving subtle spectral or morphological nuances. To address this, we tested newer YOLO variants and found YOLOv8’s computational and memory demands impractical for large-scale datasets, while YOLO11 offered a superior balance. As an optimized variant of the YOLO framework, YOLO11 retains YOLOv5’s speed while incorporating architectural advancements (e.g., dynamic convolutions, attention mechanisms) that enhance class-specific feature extraction. This enables precise detection of complex categories without exceeding our computational constraints, making it uniquely suited for resource-intensive astronomical tasks. YOLO11’s design ensures robust accuracy and efficiency, addressing both subclass discrimination and scalability challenges in astrophysical data analysis.

By integrating a simulated dataset of stars, galaxies, and quasars with YOLO11’s enhanced detection architecture, we can advance critical astrophysical classification tasks, such as: (1) star-galaxy classification, (2) galaxy redshift classification and (3) quasar searching.

\subsubsection{Star-Galaxy Classification}

\begin{figure*}[t]%
    \centering
    \begin{subfigure}
    [\centering Chip 01 (GI), ground truth]{{\includegraphics[width=0.8\columnwidth]{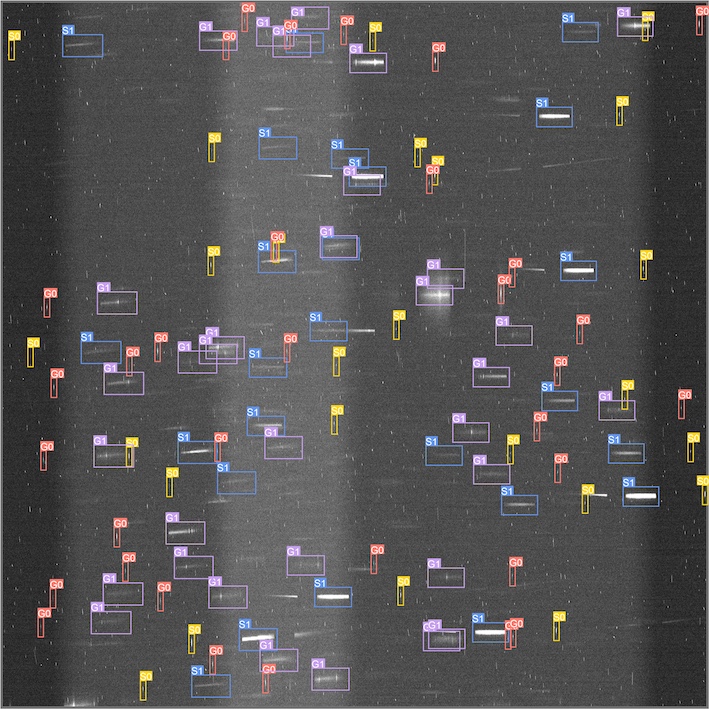}}}%
    \label{fig:YOLO11s_sg_gi_a}
    \end{subfigure}
    \qquad
    \begin{subfigure}[\centering FN vs FP]{{\includegraphics[width=0.8\columnwidth]{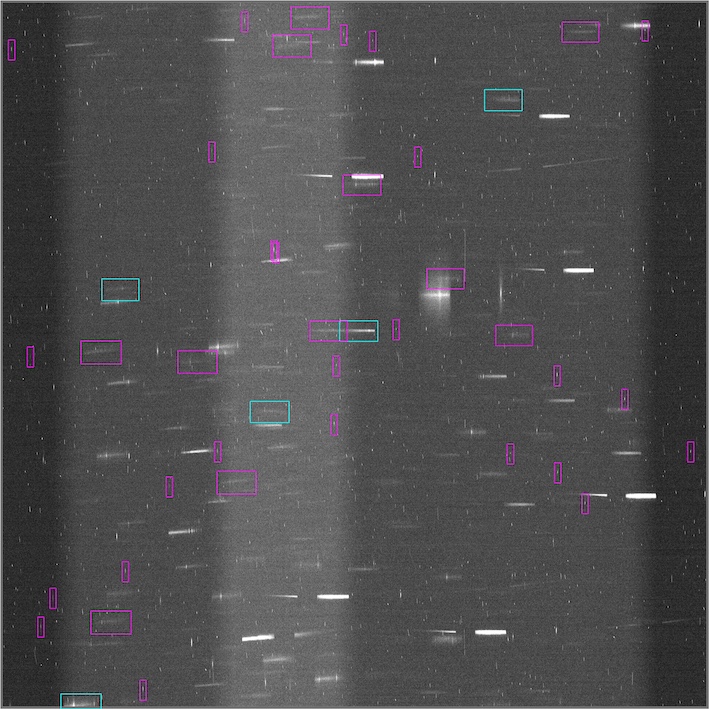}}}%
    \label{fig:YOLO11s_sg_gi_b}
    \end{subfigure}
    \\
    \begin{subfigure}[\centering Chip 01 (GI), predictions]{{\includegraphics[width=0.8\columnwidth]{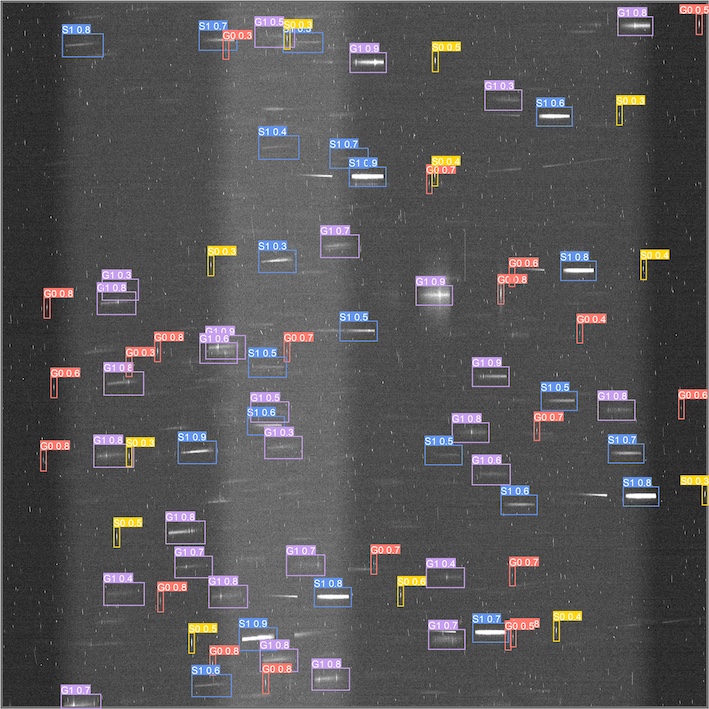} }}%
    \label{fig:YOLO11s_sg_gi_c}
    \end{subfigure}
    \qquad
    \begin{subfigure}[\centering confusion matrix]{{\includegraphics[width=0.8\columnwidth]{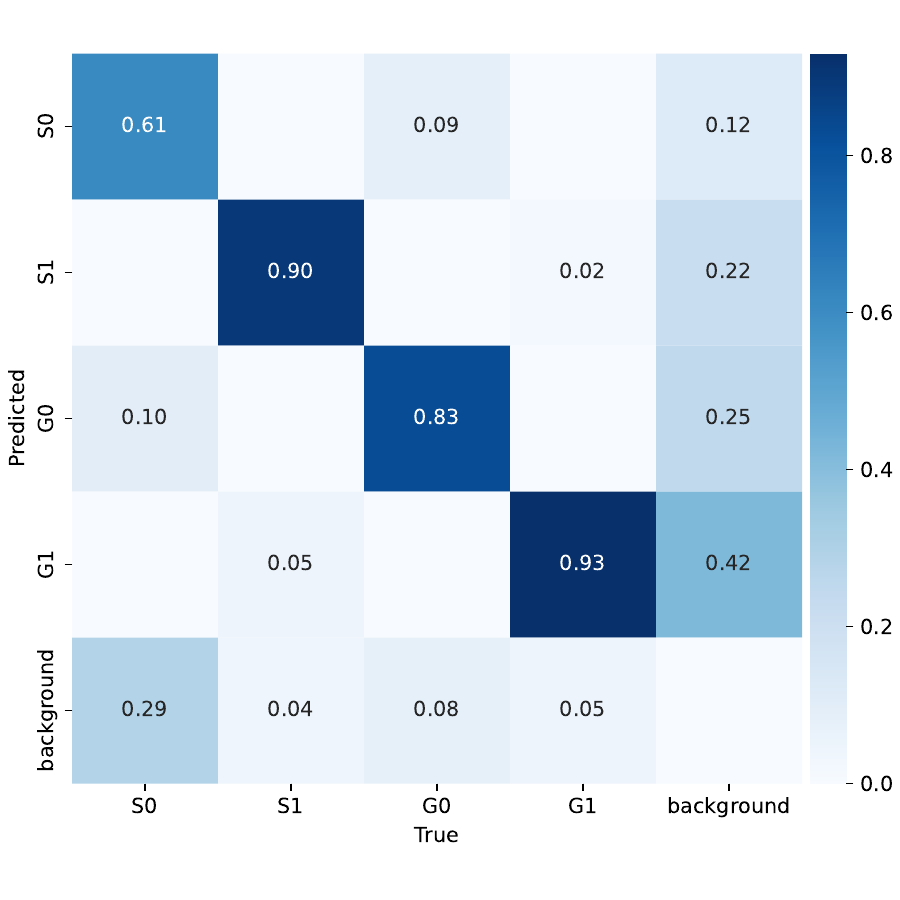}}}%
    \label{fig:YOLO11s_sg_gi_d}
    \end{subfigure}

    \caption{Performance of the YOLO11 model on star and galaxy detection. (a) Ground-truth annotations with magnitude thresholds of 23 for stars and 21 for galaxies. (b) Model predictions, with confidence scores displayed in bounding boxes. (c) Error analysis: false negatives (missed detections) highlighted in magenta and false positives (spurious detections) in cyan. (d) Confusion matrix for the test set (contains 100 sub-images), where diagonal elements represent recall rates for the four target classes.
    The results demonstrate YOLO11’s ability to discern astrophysical objects while quantifying its limitations in complex slitless spectroscopy data.}
    
    \label{fig:YOLO11s_sg_gi}
\end{figure*}

The star-galaxy classification simulation includes stars, galaxies, and quasars, though only stars and galaxies are explicitly labeled. The raw dataset comprises 20 training images, 5 validation images, and 10 test images. After cropping and resizing these images to isolate individual sources (as outlined in prior preprocessing steps), the final augmented datasets expand to 200 training samples, 50 validation samples, and 100 test samples. This scaling ensures sufficient data diversity while maintaining manageable computational loads.

Sources fainter than a defined magnitude threshold exhibit critically low SNR, hindering both model training and inference. At low SNR, the model struggles to extract meaningful spatial or spectral features from the data, leading to poor learning during training and unreliable predictions during testing. To mitigate this, magnitude thresholds are applied to exclude such low-SNR sources from the labeled dataset.

In prior work, a magnitude threshold of 22 was used for stars. However, the updated simulation features a lower source density, enabling a revised threshold strategy: galaxies are thresholded at magnitude 21, and stars at 23. This adjustment accounts for two key factors: (1)flux distribution differences: galaxies, as extended sources, exhibit dispersed flux in slitless spectral images after grating dispersion, resulting in lower per-pixel SNR compared to point-like stars of equivalent integrated magnitude; (2)
cross-band consistency: despite potential SNR variations across detector bands (e.g., differences in zero-order vs. first-order traces), thresholds are standardized (21 for galaxies, 23 for stars) to enable direct comparison of detection efficiency across bands.

The YOLO11 model’s performance on Chip 01 (GI band) is illustrated in Figure \ref{fig:YOLO11s_sg_gi}.

Figure \ref{fig:YOLO11s_sg_gi} diagnoses YOLO11’s performance on Chip 01 through a comparative analysis of ground-truth annotations (panel a) and model predictions (panel c). Panel (b) isolates detection errors: false negatives (missed faint sources, predominantly due to low SNRs) and false positives (cyan), the latter often representing dim true sources excluded by the magnitude threshold rather than misclassifications. This aligns with trends in Figure \ref{fig:fnfp}, confirming the model’s low spurious detection rate but sensitivity limitations near brightness cutoffs.

The confusion matrix (panel d) quantifies performance disparities, revealing a 90\% recall rate for first-order traces (S1, G1) but lower efficacy for zero-order detections (61\% for S0, 83\% for G0). Misclassifications remain negligible for first-order images ($\sim$5\%) but rise to $\sim$10\% for zero-order sources, reflecting the challenge of distinguishing compact stellar/galactic cores without spectral dispersion. These metrics underscore the model’s reliability for first-order analysis while pinpointing zero-order detection as a priority for refinement via SNR-aware training or adaptive thresholding.

Notably, the matrix highlights YOLO11’s precision in classifying first-order stellar/galactic spectra, attributable to their elongated morphology and distinct spectral features. Conversely, zero-order sources—lacking dispersion-axis information—exhibit greater class ambiguity, particularly for faint objects where noise dominates. This dichotomy emphasizes the need for hierarchical training strategies that weight zero-order samples more heavily or incorporate auxiliary features (e.g., spatial context from concurrent imaging).

In summary, the confusion matrix underscores the necessity of evaluating both recall (completeness) and misclassification (purity) rates to holistically assess model performance across spectral types. While YOLO11 excels in first-order analysis, zero-order detection demands targeted improvements to address SNR limitations and morphological degeneracies.

Results for the remaining bands—Chip 02 (GV) and Chip 03 (GU)—are provided in Appendix Figures \ref{fig:YOLO11s_sg_gv} and \ref{fig:YOLO11s_sg_gu}. Each chip exhibits distinct observational characteristics: variations in PSF, wavelength coverage, and sky background levels alter the dispersion geometry (length/width of spectra) and SNR for sources of equivalent magnitude. These factors directly impact detection efficacy—for instance, broader wavelength ranges or higher sky noise can reduce SNR, complicating faint source identification.

Despite lower detection rates in GV and GU bands (particularly GU) relative to GI, YOLO11 achieves consistent classification accuracy, demonstrating robustness to instrumental variations across detectors. This performance stability arises from the model’s capacity to generalize across PSF-induced trace morphology variations while maintaining discriminative power for star-galaxy separation. YOLO11's ability to maintain high classification accuracy across varied spectral conditions underscores its adaptability for analyzing multi-band slitless spectra in extensive astronomical surveys.

\subsubsection{Special Source Searching: Quasar}

\begin{figure}[h]
\centering
\includegraphics[width=\columnwidth]{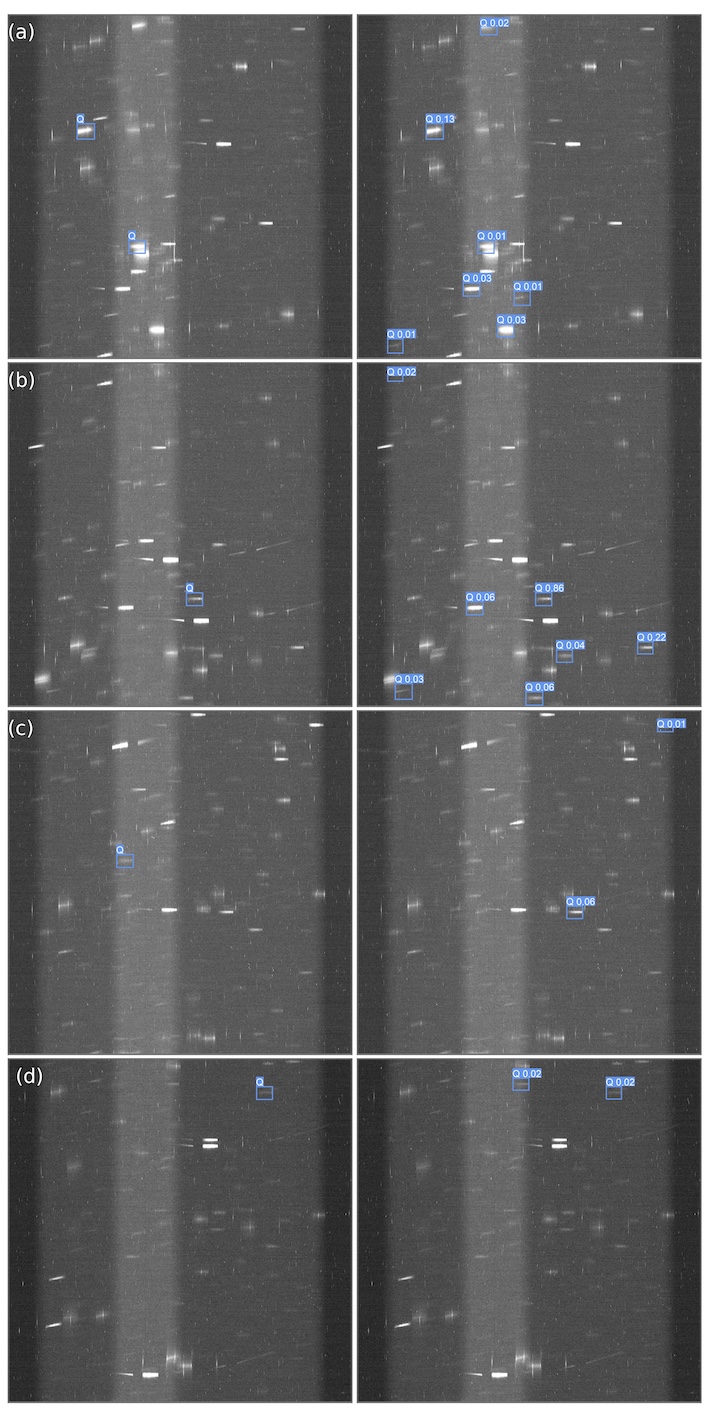}
\centering\caption{Illustrative examples of YOLO11s quasar detections in GI-band (Chip 01) at a magnitude threshold of 23.  (a) contains two bright quasars, both of which could be detected by the YOLO11s model. (b) contains a quasar with lower brightness, which could also be detected by the YOLO11s model, with the true positive having the highest score. (c) and (d) are dim sources; (c) is missed detected, while (d) is correctly detected.}

\label{fig:quasar_gt_pred}
\end{figure}

\begin{figure}[h]
\centering
\includegraphics[width=\columnwidth]{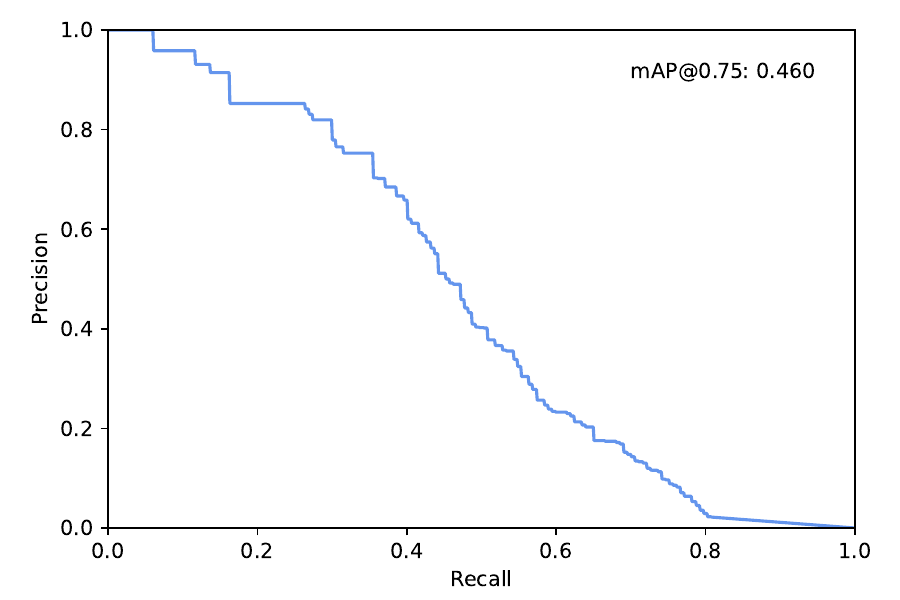} 
\centering\caption{The precision-recall curve for the YOLO11s model’s detection of first-order quasar spectra (IoU threshold: 0.75) demonstrates its ability to balance sensitivity and reliability, achieving an average precision (AP) of 0.46. AP, calculated as the area under the precision-recall curve, reflects the model’s overall capacity to localize and classify quasars across confidence thresholds. While this score indicates moderate performance, it aligns with the challenges of detecting rare, low-signal quasars in slitless spectroscopy data. To prioritize scientific completeness—critical for rare targets—we optimize for higher recall, tolerating more false positives to minimize missed detections. This ensures a comprehensive candidate list for downstream vetting, where domain-specific validation (e.g., spectral line analysis) can efficiently discard spurious detections without sacrificing potential discoveries.}
\label{fig:quasar_rp}
\end{figure}

As demonstrated in earlier analyses, the YOLO framework exhibits robust performance in identifying object positions and classifications, a capability well-aligned with the CSST's wide-field survey design. The CSST's capacity to efficiently observe vast sky regions is particularly advantageous for detecting rare, distant astrophysical phenomena such as quasars—extremely luminous active galactic nuclei (AGN) powered by supermassive black holes (SMBHs, \citealt{krolik1999agn})—represent a). Unlike quiescent SMBHs (e.g., Sagittarius A* in the Milky Way, \citealt{genzel2010galactic}), active SMBHs accrete matter at prodigious rates, ejecting relativistic jets of ionized gas that regulate star formation and galaxy evolution through 
feedback mechanisms\citep{raouf2017themany}. These jets also act as cosmic particle accelerators, offering unique laboratories for studying high-energy astrophysics.

Quasars are pivotal probes of the early universe, yet their detection remains challenging. Due to their immense distances, quasars appear as faint, point-like sources with redshifted spectra, often obscured by abundant foreground contaminants like dwarf stars\citep{caballero2008contamination}.Their low luminosity density and rarity further complicate identification in large surveys.

One application of the YOLO11 model explored in this research is the identification of quasars within CSST slitless spectral images. The model's inherent efficiency in handling large datasets and its precision in pinpointing weak signals are leveraged to address existing challenges in quasar detection. This methodology promises to improve quasar discovery yields and lay the groundwork for subsequent studies on their role in supermassive black hole (SMBH) feedback and galaxy evolution.

The dataset comprises 180 training, 20 validation, and 28 test images, each divided into 20 strips and resized to 4200 $\times$ 4200 pixels. Images lacking quasars were excluded to focus the model on relevant features. To optimize the SNR during training, we imposed a magnitude threshold of 22 for quasars in the training and validation sets, filtering out faint sources that introduce noise. Conversely, the test set includes all quasars below magnitude 23, enabling evaluation of the model’s performance on both bright and challenging faint targets. This dual-threshold strategy ensures robust learning from high-confidence examples while testing generalization to real-world conditions. The training set contains approximately 800 quasars, providing a foundation for the model to discern quasar-specific spectral signatures amidst contaminants like dwarf stars. 

We present four examples to compare the ground truth and predictions in Figure \ref{fig:quasar_gt_pred}. The ground truth are shown in the left column, while the predictions with corresponding scores (or confidence levels) are displayed in the right column.
(a) contains two bright quasars. The YOLO11s model detected them correctly, even though one quasar overlaps with other bright sources; (b) features a less bright quasar, which the YOLO11s model successfully detected with the highest confidence score; (c) is an example of a missed detection, where the quasar has a high magnitude, making it more challenging to identify; (d) demonstrates a successful detection of a high-magnitude quasar, showcasing the model's capability to identify faint sources accurately.

The YOLO11s model demonstrates strong capability in detecting quasars across varying brightness levels, excelling at identifying bright quasars even in crowded fields with overlapping sources (Figure \ref{fig:quasar_gt_pred}).
However, the model struggles with faint (high-magnitude) quasars and overlapping sources, where low signal-to-noise ratios (SNRs) lead to missed detections. Despite these challenges, the model’s ability to recover some faint quasars—albeit with lower confidence—highlights its potential for advancing quasar detection in slitless spectral surveys, particularly when supplemented by secondary validation methods to filter low-confidence candidates.

The inherent trade-off between recall (completeness) and precision (accuracy) is evident in the model’s precision-recall curve (Figure \ref{fig:quasar_rp}), which achieves an AP of 0.46 under strict localization criteria (IoU > 0.75). While the AP score of 0.46 reflects moderate effectiveness, this aligns with the inherent difficulty of identifying rare, faint quasars in slitless spectroscopy data, where low SNRs and spectral blending pose significant hurdles. To ensure scientific completeness for rare targets like quasars, we prioritize maximizing recall—even at the cost of increased false positives—to reduce missed detections. This strategy produces an inclusive candidate pool that subsequently undergoes secondary validation using PyQSOFit \citep{PyQSOFit2018}, a tool that performs spectral line analysis to efficiently eliminate false alarms and further decompose the components of potential discoveries. The AP score 0.46 underscores the model’s value as a first-pass detection tool in large-scale surveys like the CSST, where rapid, broad-scope candidate identification is critical, even as refinements remain necessary to enhance precision for faint or overlapping sources.

\subsubsection{Galaxy Redshift}

\begin{figure*}[t]%
    \centering
    \begin{subfigure}[\centering Chip 01 (GI), galaxy redshift, ground truth]{{\includegraphics[height=0.6\columnwidth]{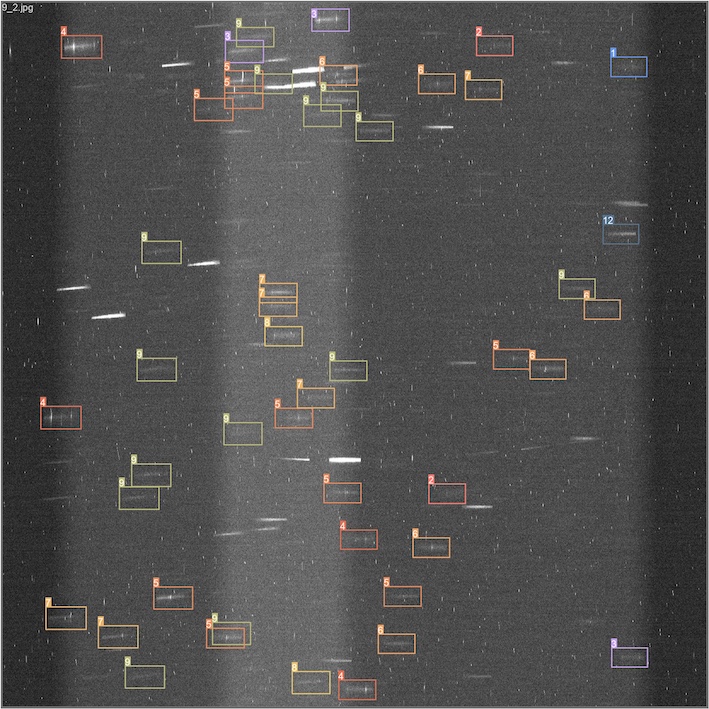}}}%
    \label{fig:galaxy_z_a}
    \end{subfigure}
     ~~
    \begin{subfigure}[\centering Chip 01 (GI), galaxy redshift, predictions]{{\includegraphics[height=0.6\columnwidth]{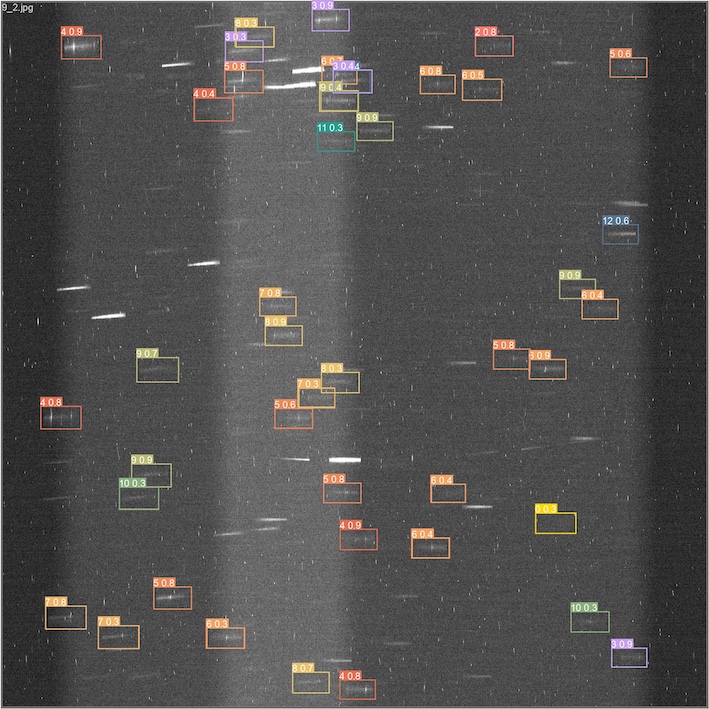}}}%
    \label{fig:galaxy_z_b}
    \end{subfigure}
    ~~
    \begin{subfigure}[\centering confusion matrix]{\includegraphics[height=0.6\columnwidth]{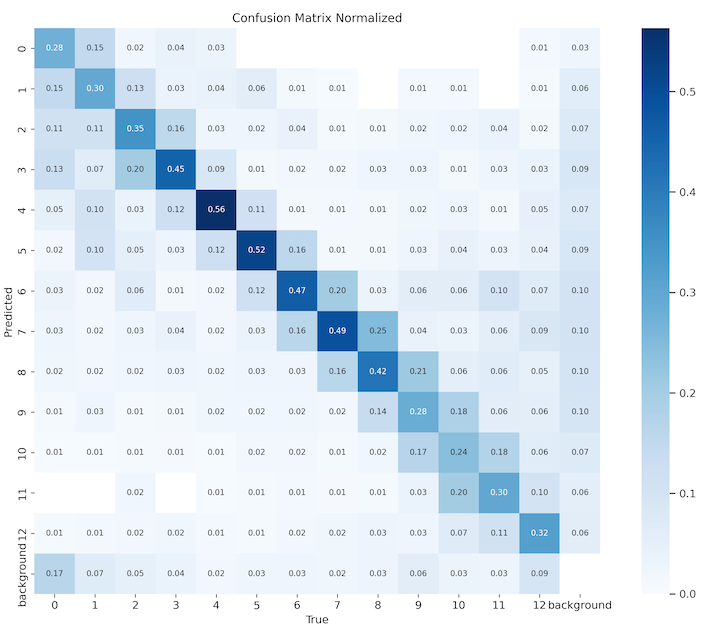}}%
    \label{fig:galaxy_z_c}
    \end{subfigure}

    \caption{Galaxy redshift classification using the YOLO11 model. (a) Ground-truth annotations with galaxies binned into 13 redshift categories. (b) Model predictions showing detected redshift bins and associated confidence scores. (c) Confusion matrix evaluating classification accuracy, where diagonal elements reflect correct predictions. A magnitude threshold of 21.5 was applied to galaxies in the dataset to ensure signal-to-noise reliability.}

    \label{fig:galaxy_z}
\end{figure*}

Understanding and measuring the redshift of galaxies is fundamental to comprehending the universe's structure, history, and dynamics. Redshift enables astronomers to gauge the distance to galaxies, as the light from these distant objects is stretched to longer wavelengths due to the universe's expansion. This phenomenon not only provides compelling evidence for the expanding universe but also aids in estimating the Hubble constant, which quantifies the rate of this expansion. Moreover, redshift data is invaluable for studying the formation and evolution of galaxies by allowing us to observe them as they were in the distant past. It also plays a crucial role in mapping the large-scale structure of the universe, revealing the intricate distribution of galaxies, clusters, and voids. Additionally, redshift measurements are pivotal in investigating dark energy, the enigmatic force driving the accelerated expansion of the universe. Overall, redshift serves as a powerful tool that enhances our understanding of the cosmos and helps validate and refine cosmological models.

Methods for measuring redshift range from established spectroscopic line analysis \citep{colless2001spec_redshift} and photometric template fitting \citep{benitez2000pho_redshift} to contemporary machine learning techniques that can be applied to both photometric and spectroscopic data \citep{firth2003neu_redshift, hoyle2016deep_redshift, kind2013random_forest_redshift, Zhong2024gasnet}. While low resolution inherently limits the precision of any redshift determination method, machine learning offers a powerful alternative to traditional spectral fitting by focusing on pattern recognition in the overall spectral shape and statistical relationships, making it more robust and potentially more accurate for analyzing large datasets of low-resolution galaxy spectra.

Traditional YOLO models are designed for discrete object classification (like distinguishing stars from galaxies). However, predicting galaxy redshift is a continuous regression problem. To utilize YOLO for this task, we've adopted a binning strategy, transforming the regression into a multi-class classification problem. This initial approach allows us to leverage YOLO's strengths to explore its potential in redshift estimation, acknowledging inherent limitations in precision and potential ambiguities at bin edges. Although binning reduces the redshift granularity, it provides a practical balance between simplicity and usefulness for this initial investigation. Our subsequent work will involve modifying YOLO's architecture to directly predict continuous redshift values, aiming for higher precision needed in detailed spectral analysis. This phased approach enables us to validate the model's core capabilities while laying the groundwork for more sophisticated, regression-based frameworks.

We constructed a dataset of 60 slitless spectral images (30 training, 10 validation, 20 test). Each image underwent processing involving division into 10 strips and resizing to 4200 $\times$ 4200 pixels to isolate individual sources. Our analysis concentrates on first-order galaxy spectra brighter than magnitude 21.5, a threshold chosen to ensure sufficient SNR, given that most galaxies in this range have redshifts under 1.2. To adapt the YOLO model for redshift prediction, we implemented a binning strategy, discretizing continuous redshift values into 13 bins of width 0.1, with the final bin encompassing redshifts $>$1.2. This binning scheme balances the need for redshift granularity with computational efficiency.

The results are presented in Figure \ref{fig:galaxy_z}, where panel (a) shows ground-truth redshift bins, (b) displays model predictions with confidence scores, and (c) presents a confusion matrix evaluating classification accuracy across bins. This structured approach demonstrates the model’s capability to categorize redshifts while highlighting challenges near bin boundaries, laying the groundwork for future regression-focused adaptations. 

The confusion matrix reveals strong diagonal dominance, indicating the model’s ability to correctly classify most galaxies into their redshift bins. However, adjacent bin misclassifications occur due to the inherently continuous nature of redshift, where subtle spectral differences near bin boundaries challenge discrete categorization. Class imbalance further impacts performance: bins with abundant samples (e.g., classes 4–8) achieve higher accuracy, while sparsely populated bins (e.g., classes 1, 11) suffer from lower precision, reflecting the model’s reliance on data volume for robust learning. To mitigate this, techniques like data augmentation, class-weighted loss functions, or resampling could balance representation across bins. 

Even with a finer granularity of 25 redshift bins, the confusion matrix presented in Figure \ref{fig:YOLO11s_galaxy_redshift_24} of Appendix \ref{sec:appendix_figure} confirms the robustness of our conclusion.

To overcome the limitations of discrete classification, including binning artifacts and reduced precision (particularly for high-redshift galaxies), we aim to transition the YOLO framework to direct continuous redshift regression. This will involve integrating regression-specific adaptations, such as modified output layers and loss functions, with the goal of maintaining YOLO’s detection efficiency while achieving spectroscopic-level accuracy. Coupled with class-balancing strategies, these advancements will position the model as a scalable tool for precision cosmology in slitless spectroscopy surveys.

\section{Discussion}
\label{sec:discussion}

\subsection{Model Performance}

The proposed deep learning framework demonstrates significant potential in addressing the unique challenges of detecting and classifying slitless spectroscopic data from the CSST. By adapting YOLO-based architectures, the model achieves robust performance across varied stellar densities and spectral bands. On validation datasets, YOLOv5 attains a mean average precision (mAP) of 0.966 (IoU=0.7) for detecting zero- and first-order spectra, even in crowded fields where spectral image overlapping with each other. 

YOLO11 classification extends to astrophysical object typing, achieving 90\% recall for distinguishing stellar and galactic first-order spectra. The model also demonstrates capability in detecting rare, low-SNR astrophysical sources such as quasars and providing coarse-grained redshift estimates for galaxies via classification into discrete bins.

\subsection{Implicit Deblending of Spectra}

\begin{figure*}[h]%
    \centering
    \begin{subfigure}[\centering spectral image]{{\includegraphics[width=0.6\columnwidth]{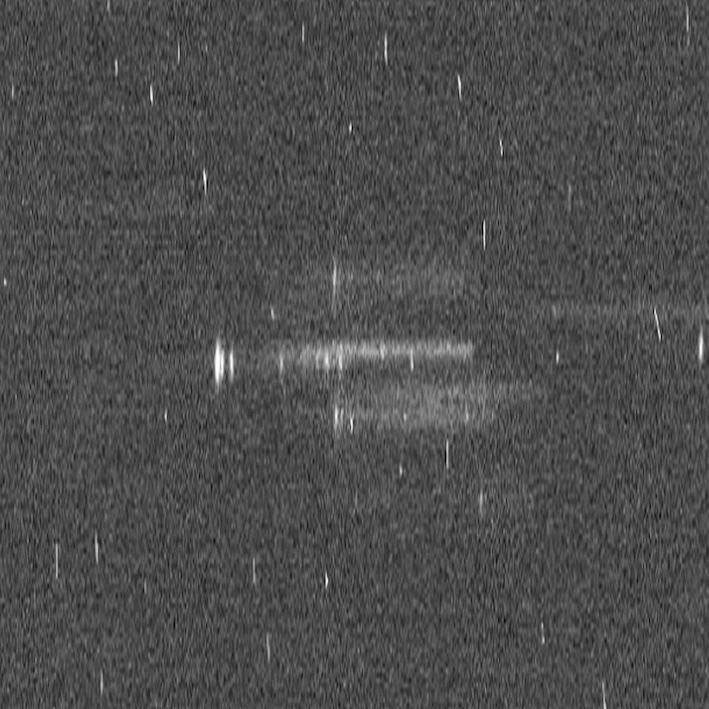} }}%
    \label{}
    \end{subfigure}
   ~~
    \begin{subfigure}[\centering ground truth]{{\includegraphics[width=0.6\columnwidth]{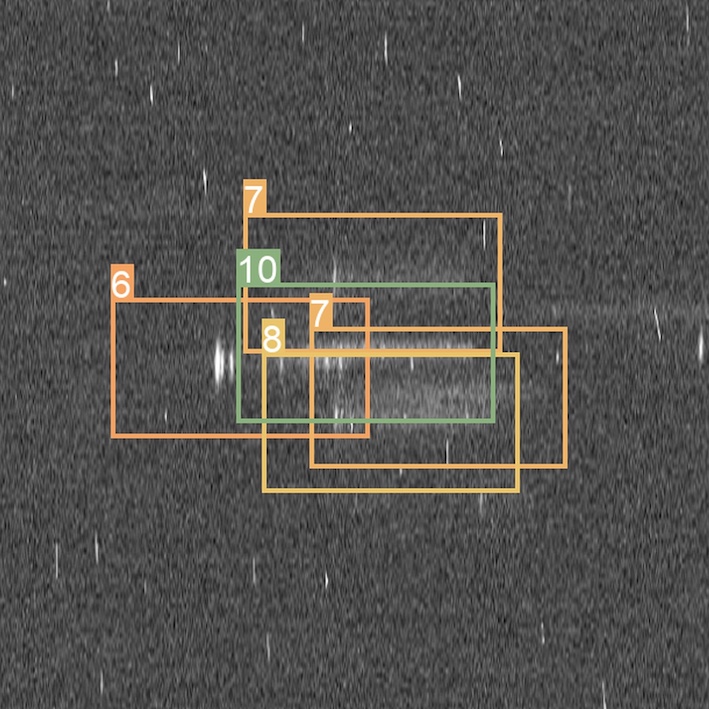}}}%
    \label{}
    \end{subfigure}
   ~~
    \begin{subfigure}[\centering prediction]{{\includegraphics[width=0.6\columnwidth]{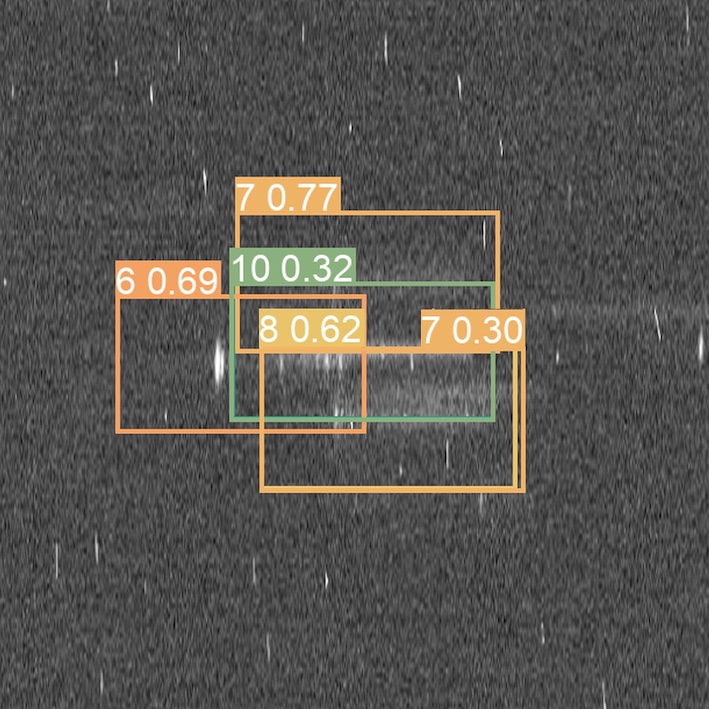} }}%
    \label{}
    \end{subfigure}
    \qquad
    \caption{ Performance of the YOLO11s model in a challenging case of overlapping spectra. (a) The composite image showing five blended first-order galaxy spectra. (b) Ground-truth annotations for the five galaxies, corresponding to redshift bins 6, 7, 7, 8, and 10. (c) Model predictions. Despite significant spectral blending, the model successfully detects and correctly classifies all five sources. This demonstrates its ability to deconvolve complex scenes, with only a minor positional offset visible for one of the traces in bin 7.}
    
    \label{fig:dense_crop}
\end{figure*}

A key advantage of our deep learning approach is its ability to handle overlapping spectra implicitly, removing the need for a dedicated deblending module. When a bounding box contains contaminating flux from nearby sources, the model is trained to isolate the primary target, effectively treating the blended light as complex background noise; Figure \ref{fig:dense_crop} illustrates this process with selected successful examples from Simulation B. This approach stands in contrast to traditional data processing pipelines, which must handle overlaps by either excluding the blended regions entirely—thereby losing valuable data—or by attempting to explicitly model and subtract the contamination. Our method proves highly effective, with current limitations in extremely dense fields (like dense field mwl90b10 in Simulation A) being computational (CUDA memory) rather than algorithmic.

\subsection{Bounding Box Offset}

\begin{figure*}[t]
\centering
\includegraphics[width=1.8\columnwidth]{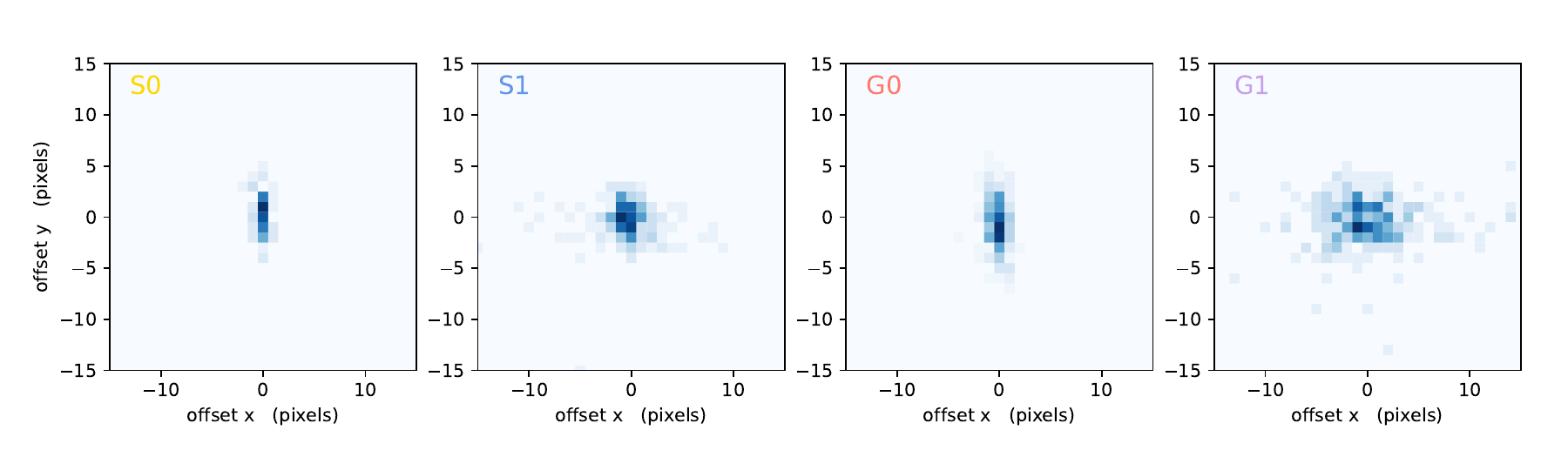}
\centering\caption{Bounding box offsets for star-galaxy classification in images of size 4224$\times$4224. The offsets in the y-axis are around 5 pixels, while in the x-axis, the offsets are approximately 1, 3, 2, and 5 pixels for S0, S1, G0, and G1, respectively. After rescaling to the original image size, the offsets in the y-axis are around 1 pixel, and in the x-axis, they are approximately 2, 6, 4, and 10 pixels for S0, S1, G0, and G1.}
\label{fig:bounding_box_offsets}
\end{figure*}

Accurate spatial localization via bounding boxes is essential for pinpointing astrophysical targets in slitless spectroscopy. While the YOLO model delivers robust object detection and localization, the bounding boxes it generates do not inherently encode spectral information, such as the wavelength corresponding to each pixel. This gap precludes direct derivation of the wavelength-flux relationship (i.e., 1D spectra) from YOLO’s output alone—a critical limitation for spectroscopic analysis.

In the CSST pipeline, wavelength calibration relies on correlating detected spectral traces with their pre-dispersion source positions (or zero-order trace positions). Consequently, quantifying positional offsets of the centers between predicted and ground-truth bounding boxes of the zero-order image becomes vital. These offsets, which reflect systematic errors in the model’s localization accuracy, enable post-detection corrections to align spectral traces with their true spatial and spectral coordinates. Figure \ref{fig:bounding_box_offsets} analyzes the spatial offsets identified by the YOLO11s model, comparing detection accuracy between critical zero-order calibration classes (S0, G0) and supplementary first-order detections (S1, G1). These offsets were evaluated using a 0.7 intersection-over-union (IoU) threshold to align predicted and ground-truth bounding boxes. The results highlight distinct class-specific spatial variations, enabling precision adjustments to refine wavelength calibrations and subsequent spectral extraction workflows.

Following image resizing from the original 9216 $\times$ 979 pixels to 4224 $\times$ 4224 pixels, scale factors of $\sim$2.18 (x-axis) and $\sim$0.23 (y-axis) were applied to map offsets between resized and original dimensions. Y-axis offsets for all four classes (S0, G0, S1, G1) average $\sim$1 pixel in resized coordinates (equivalent to 5 $\times$ 0.23  $\approx$ 1 pixel original scale). X-axis offsets, however, vary significantly by class: S0 exhibits the smallest resized offsets (0–1 pixels, translating to $\sim$2 original pixels), followed by G0 (0–2 resized pixels, $<$4 original pixels), S1 ($\sim$3 resized pixels, $\sim$6 original pixels), and G1 (up to 5 resized pixels, $\sim$10 original pixels). This progression highlights increasing x-axis spatial discrepancies from S0 to G1, with G1 detections showing the largest scaling-induced errors, informing prioritization for calibration refinement.

The spatial offsets of our zero-order detections are small, therefore our detection method offers a direct solution to the problem of zero-order trace localization for slitless spectral wavelength calibration. The YOLO based detection method provides accurate and immediate identification of these positions from the slitless images, eliminating dependence on time-delayed multicolor imaging positions matching and incomplete external star catalogs for CSST's deep observations, leading to a robust and timely self-contained calibration.

\subsection{Detection Limits - Magnitude}

\begin{figure*}[t]%
    \centering
    \begin{subfigure}[\centering false negatives and false positives in star-galaxy classification, Chip 01 (GI)]{{\includegraphics[width=0.8\columnwidth]{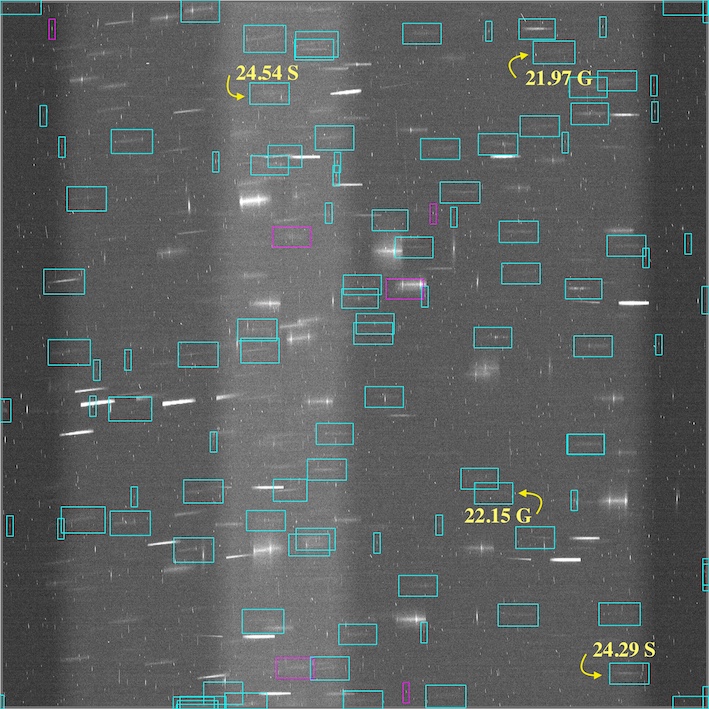}}}%
    \end{subfigure}
     ~~
    \begin{subfigure}[\centering True boxes, with star magnitude from 23 to 25, galaxy magnitude 21 to 22.2]{{\includegraphics[width=0.8\columnwidth]{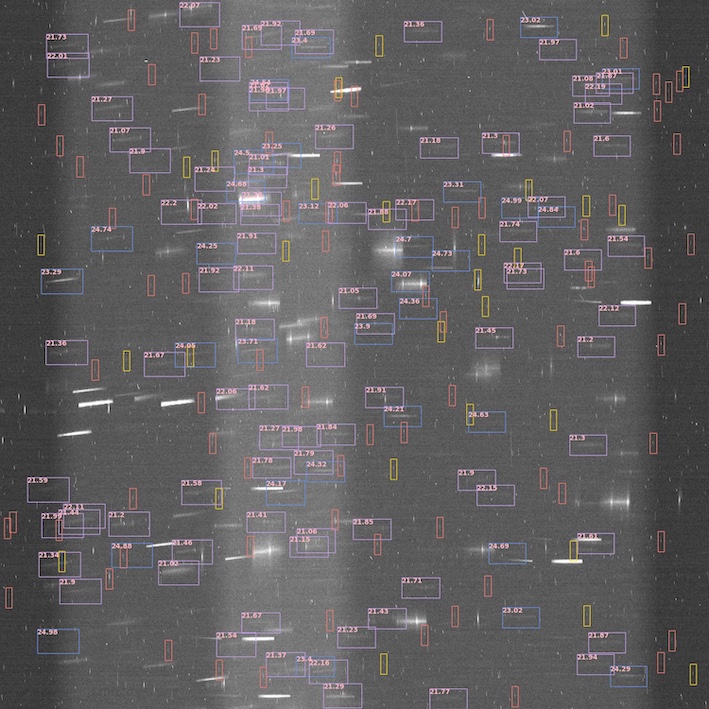}}}%
    \end{subfigure}

    \caption{(a) Illustrates the false positives (cyan) and false negatives (magenta) in the YOLO11 predictions. (b) Depicts the bounding boxes of faint sources that exceed the magnitude threshold, with magnitudes ranging from 23 to 25 for stars and 21 to 22.5 for galaxies. Cross-referencing both panels reveals the faintest detected sources—a star at  and a galaxy at —as explicitly labeled in panel (a). This demonstrates the model's capability to detect faint objects beyond the predefined magnitude threshold, underscoring its potential for precise astronomical analysis.}
    
    \label{fig:mag_limit}
\end{figure*}

Our analysis shows that most false negatives arise from astrophysical sources intentionally excluded from ground-truth labels (see Figure \ref{fig:fnfp}) due to magnitude thresholds designed prevent bounding box overlaps and ensure sufficient signal-to-noise ratios for reliable labeling. This reveals YOLO11’s ability to detect sources below these operational brightness limits (stars: m=23, galaxies: m=21).

To understand the detection limits, an example is illustrated in Figure \ref{fig:mag_limit}. The left panel (a) displays YOLO11's prediction errors, with false positives shown in cyan and false negatives in magenta. Cross-referencing these false positives with cataloged faint sources we found that most of the false positives align with genuine faint sources, stars at magnitude 23–25 and galaxies at magnitude 21–22.5. The figure explicitly labels the faintest detected sources in this image, with stellar magnitudes reaching 24.54 and galactic systems detected down to 22.15\footnote{They are exemplifies, there are still a lot of sources brighter could not be detected}. 

This observation highlights the impressive sensitivity of the YOLO11 model, which can detect objects beyond the predefined magnitude threshold, up to 1.5 magnitude for stars and 1.1 for galaxies. It also underscores the importance of considering faint sources in the analysis, as they contribute valuable information to the overall dataset. By refining the magnitude threshold and incorporating these faint sources, we can further enhance the accuracy and robustness of the YOLO11 model in detecting and classifying slitless spectral images.

The scientific utility of our framework is ultimately defined by its detection limits and completeness, as these metrics determine its applicability to key CSST science objectives. Our analysis reveals that the model achieves a high degree of completeness, with a recall rate exceeding 90\% for first-order spectral traces down to magnitude 23 for stars and magnitude 21 for galaxies. This robust performance ensures the construction of vast, statistically significant catalogs essential for foundational CSST science programs. For instance, these capabilities are exceptionally well-suited for Galactic archaeology, which relies on large, unbiased samples of stars to map the Milky Way's structure and chemical history. Similarly, our framework can generate the comprehensive galaxy catalogs required for mapping the large-scale structure of the cosmos.

Beyond the high-completeness regime, our model demonstrates impressive depth, with the validated detection of individual stellar sources as faint as magnitude 24.5. This ability to probe deeper than our conservative completeness limits highlights the framework's significant potential for new discoveries. Nevertheless, we acknowledge that the most demanding of CSST's science cases, particularly precision weak lensing studies, require pushing the boundaries of detection to even fainter galaxies to achieve the necessary source density for robust cosmological constraints. Therefore, while our current framework represents a critical advancement, future work will focus on targeted enhancements—such as refined loss functions and advanced data augmentation—to improve sensitivity at the faintest detection limits. These strategic improvements will ensure our framework can maximally exploit the rich dataset across all facets of the CSST survey, from Galactic studies to precision cosmology.

\subsection{End-to-end Workflows}
CSST faces two critical challenges: (1) spatial uncertainties caused by non-simultaneous imaging, which introduce positional inaccuracies in crowded fields, and (2) spectral blending, where overlapping traces hinder the isolation of individual sources. To address first issue, we leverage YOLO models to detect and spectral traces directly from raw slitless images.

YOLO excels not only in detecting and localizing spectral traces under challenging conditions but also in extracting scientifically meaningful features, which may help us to solve problem (2). 

The YOLO11 model distinguishes between stellar and galactic spectra with high precision, enabling applications such as quasar detection and coarse-grained redshift estimation (providing discrete labels like ). Although currently, the YOLO model cannot provide galaxy redshifts with the same accuracy as traditional methods—it is far exceeding the precision of traditional template-matching pipelines ($\Delta$z$<$0.005)—it shown that the feature of sources could be learned from the 2D spectral image. 

Our work establishes the foundation for a unified deep learning framework that automates astrophysical parameter extraction from slitless spectral images. While current YOLO implementations cannot yet match the accuracy of specialized fitting methods for parameters like galaxy redshifts, they demonstrate the feasibility of end-to-end learning: the model processes raw images and outputs parameters directly, eliminating manual steps (e.g., background subtraction, wavelength calibration) and their associated errors. This approach minimizes error propagation, accelerates analysis, and leverages machine learning’s ability to uncover subtle, non-linear relationships in the data. 

This end-to-end approach offers several advantages. It simplifies the workflow by eliminating intermediate steps, which can be time-consuming and error-prone. By directly processing raw images, the model streamlines analysis and provides faster results. Additionally, it integrates advanced machine learning techniques that can learn complex patterns and relationships within the data, potentially leading to more accurate and robust parameter estimations.

Moreover, the end-to-end approach aligns with the trend towards automation and efficiency in astronomical data analysis. As the volume of data from telescopes like the CSST increases, the ability to process and analyze this data quickly and accurately becomes crucial. End-to-end models like YOLO enable astronomers to handle large datasets more effectively and focus on interpreting results rather than managing data processing.

In summary, while the YOLO model may not yet match the accuracy of traditional methods for determining parameters such as galaxy redshifts, its end-to-end approach offers significant potential for streamlining and enhancing the analysis of slitless spectral images. As machine learning techniques evolve, we can expect further improvements in the accuracy and reliability of these models, making them valuable tools for astronomical research.

\subsection{From Simulation to Reality}

While our model demonstrates strong performance, it is important to acknowledge that it has been trained exclusively on simulated data. As a result, the model might not generalize directly to real CSST data without further adaptation. After we obtain a partially labeled set of real observational data, we can employ transfer learning techniques to bridge this gap. This would involve freezing the weights of the early layers of the network, which have learned general, low-level features, and then fine-tuning the deeper layers using the new, real-world data.

This approach is practical because, in terms of data characteristics, there is a reasonable expectation of transferability. Although the errors and noise in real data may be larger and more complex than in the simulation (as some effects are impossible to perfectly model), the fundamental features of the targets should remain consistent. Moreover, real observational data may contain additional, undiscovered features not present in the simulations, and fine-tuning will allow the model to learn and adapt to this richer, more complex data environment.

\section{Conclusion} 
\label{sec:conclusion}

In conclusion, our study validates the potential of YOLO models for detecting and classifying slitless spectral traces in CSST data. We demonstrate that YOLOv5 achieves robust detection of zero- and first-order spectral images across fields of varying source densities, including crowded regions where overlapping spectra dominate. In medium-density fields, YOLO11 further distinguishes itself by learning class-specific features of slitless spectral traces, enabling precise localization and classification of astronomical objects. This capability not only facilitates differentiation between stars and galaxies but also empowers targeted searches for rare or unique sources, such as quasars, within slitless spectral datasets.

YOLO11 also demonstrates its utility in determining galaxy redshifts (coarse-grained), a critical task for measuring cosmic distances. Unlike traditional pipelines reliant on multi-stage spectral analysis, YOLO11 achieves this through an end-to-end workflow, processing raw slitless images to deliver redshift estimates without intermediate manual steps.

Our model achieves minimal offsets between predicted and ground-truth bounding boxes, demonstrating high localization accuracy. Along the y-axis, positional deviations are as small as 1 pixel, while for zero-order spectral traces on the x-axis, offsets remain below 2 pixels. This sub-pixel precision in zero-order trace localization enables robust wavelength calibration—a cornerstone of reliable spectral analysis—by minimizing spatial uncertainties that could propagate into systematic errors in derived astrophysical parameters.

Our experiments underscore the critical importance of tailoring model selection—version, size, and configuration—to computational constraints and scientific objectives. YOLOv5, with its streamlined architecture and rapid inference speed, serves as a pragmatic choice for real-time applications or dense stellar fields where processing efficiency is paramount. Conversely, YOLO11’s architectural sophistication and enhanced feature-learning capabilities make it better suited for intricate tasks such as star-galaxy classification and quasar detection, where accuracy outweighs computational overhead. This tiered approach ensures optimal performance across diverse observational scenarios, balancing speed and precision to meet the demands of modern slitless spectroscopy.

Future work will focus on overcoming current limitations, including advancing the model’s capacity to handle continuous astrophysical variables (e.g., galaxy redshifts, stellar metallicities) through refined regression architectures. Additionally, we aim to enhance sensitivity to faint objects in low signal-to-noise regimes by optimizing feature extraction layers and loss functions. We also plan to integrate other deep learning techniques to further improve accuracy and robustness.

Our findings underscore the potential of YOLO models in advancing astronomical image analysis, allowing for more accurate and efficient processing of large datasets. By integrating YOLO models into slitless spectral image analysis, we significantly enhance our ability to process and interpret complex astronomical data, paving the way for future discoveries and innovations in astronomy.

\section{Acknowledgments}

This research is supported by the Beijing Natural Science Foundation Youth Program (No. 1244061), the science research grants from the China Manned Space Project (No. CMS-CSST-2025-A11), the National Key Research and Development Program of China (No. 2022YFF0504200), and China Manned Space Program through its Space Application System. H.T. is supported by the National Key R\&D Program of China No. 2024YFA1611902.
The authors thank the foundation for its support of astrophysical data analysis methodologies. N.L. acknowledge the support from the Ministry of Science and Technology of China (No. 2020SKA0110100), the science research grants from the China Manned Space Project (No. CMS-CSST-2021-A01) and CAS Project for Young Scientists in Basic Research (No. YSBR-062).

The authors acknowledge the use of \textsc{DeepSeek} (DeepSeek Inc.)\footnote{\url{www.deepseek.com}} for refining the language and structure of this manuscript. The tool assisted in improving readability and coherence, but all scientific content, interpretations, and conclusions remain the sole responsibility of the authors.

\appendix 
\renewcommand{\thefigure}{A\arabic{figure}}
\setcounter{figure}{0}

\section{YOLO}
\label{sec:YOLO}
\subsection{A Brief History of YOLO}
The YOLO family of models has profoundly shaped the field of object detection by pioneering real-time, end-to-end frameworks. Introduced in  \cite{redmon2016you}, YOLO became the defining model for one-stage detection, by prioritizing rapid inference through single-pass prediction of bounding boxes and class probabilities directly. However, this approach initially traded some localization precision, as evidenced by higher errors compared to the contemporaneous Fast R-CNN \citep{girshickICCV15fastrcnn}. The release of YOLOv2 \citep{redmon2017YOLO9000} marked a leap forward, integrating anchor boxes for improved bounding box alignment, batch normalization for stable training, and the streamlined Darknet-19 backbone, collectively enhancing both accuracy and speed. Building on this, YOLOv3 \citep{redmon2018YOLOv3} incorporated feature pyramid networks (FPN, \citealt{lin2017fpn}) into its architecture, enabling multi-scale predictions. By detecting objects at three distinct scales, YOLOv3 effectively mitigated prior limitations in identifying small objects, a notable weakness in YOLOv1 and YOLOv2. These iterative advancements underscored YOLO’s evolution toward balancing speed, precision, and versatility in diverse detection scenarios.

YOLOv4, introduced in \cite{bochkovskiy2020YOLOv4}, marked a methodological leap in object detection by systematically integrating two novel concepts: "Bag of Freebies" (BoF) and "Bag of Specials" (BoS). BoF encompasses training optimizations—such as advanced data augmentation, learning rate scheduling, and loss function refinements—that enhance model accuracy without inflating inference time. In contrast, BoS introduces lightweight architectural enhancements, including attention mechanisms and post-processing modules, which marginally increase training complexity but significantly boost detection precision.

A key advancement in YOLOv4 was its improved capability for small object detection. By adopting higher input resolutions, the model expanded its network depth (to capture broader contextual receptive fields) and parameter count (to enhance representational capacity). This necessitated a backbone architecture optimized for both efficiency and performance, leading to the selection of CSPDarknet53 through rigorous experimentation. Its cross-stage partial connections reduce computational redundancy while maintaining feature richness, making it ideal for high-resolution detection tasks. For a comprehensive breakdown of BoF/BoS components and architectural trade-offs, see \cite{bochkovskiy2020YOLOv4}.

YOLOv5 extends the advancements of YOLOv4 by transitioning to a PyTorch framework, enhancing accessibility and integration with modern machine learning workflows. Key innovations include the AutoAnchor algorithm, which dynamically optimizes anchor box dimensions during training to align with dataset-specific object scales, and computational efficiency tailored for high-resolution images. 

Subsequent YOLO variants, comprehensively reviewed in \citep{terven2023comprehensive}, demonstrate continual evolution in object detection. However, publication timelines do not inherently correlate with model efficacy for domain-specific challenges like our ultra-high-resolution slitless spectra. Prioritizing the trade-off between speed and computational feasibility, we evaluated multiple architectures. Through comprehensive evaluation, YOLOv5 and YOLO11 were selected as complementary solutions: YOLOv5 works well in crowded star regions where other models crash-- due to computational overload or memory constraints; YOLO11, however, prioritizes precision and granularity, which enables astrophysical parameter extraction in the future.

\subsection{The Construction of YOLO}

\subsubsection{Building Blocks}

To understand how YOLOv5 works, we start by breaking down its architecture into core components. Though initially complex, its design is built from interconnected elements. Understanding these components forms the basis for grasping its full architecture.

ConvBNSiLU: The ConvBNSiLU module in YOLOv5 integrates three core components for efficient feature extraction: a convolution layer (Conv) that applies learnable filters to detect spatial patterns (edges, textures) by sliding kernels over input data, batch normalization (BN) to standardize feature map outputs (mean $\sim$ 0, variance $\sim$ 1), stabilizing training and reducing overfitting, and the Sigmoid Linear Unit (SiLU) activation function, which introduces non-linearity via ${\rm SiLU}(x)=x\sigma(x)$, which retains ReLU’s gradient-preserving efficiency for positive inputs while incorporating sigmoid’s smoothness to handle negative inputs gracefully. Together, these components enable robust hierarchical feature learning, balancing computational efficiency with the precision required for tasks like detecting faint spectral traces or resolving overlapping objects in high-resolution slitless data.

C3: The C3 module (Cross Stage Partial Bottleneck with 3 convolutions) in YOLOv5 optimizes feature extraction by combining cross-stage partial connections \citep{wang2019} with a bottleneck structure inspired by ResNet \citep{he2015}. Designed to reduce computational complexity and memory demands, it addresses challenges in training deep networks by compressing features through sequential convolutions while preserving performance. Cross-stage connections enable efficient gradient flow and multi-scale feature integration by linking network blocks, enhancing information propagation. 
The C3 module processes input data through two parallel pathways. Initially, both the main path and the secondary path undergo ConvBNSiLU layers for initial feature extraction. The main path then proceeds through a bottleneck layer. Meanwhile, the secondary path bypasses the bottleneck, preserving raw feature details. Finally, the outputs of both paths are concatenated to merge compressed high-level features with retained low-level details, followed by a final ConvBNSiLU layer to refine the fused features. This dual-path design optimizes efficiency and feature richness, enabling precise detection of complex patterns in high-resolution data.

SPPF: SPP stands for spatial pyramid pooling, which was introduced to deep convolutional networks for visual recognition by \citet{He_2014}; F stands for fast. The SPP layer pools the features and generates fixed-length outputs, which removes the fixed-size constraint of a CNN. 

\subsubsection{Architecture} 

Modern neural networks for computer vision are typically structured into three core components: a feature extraction backbone that progressively encodes hierarchical patterns from raw pixels to semantic concepts, a multi-scale fusion neck that bridges coarse and fine-grained features to resolve ambiguities in object size or occlusion, and a task-specific head that translates enriched features into predictions.

The backbone plays a critical role in transforming input data into a meaningful feature representation. Functioning as the core feature extractor, this component progressively identifies layered patterns within the input data—beginning with basic elements like edges and textures in initial layers, then advancing to recognize complex structures and contextual relationships in deeper network stages. This hierarchical processing enables the model to resolve intricate patterns, such as overlapping spectral traces, by progressively integrating low-level features (e.g., edges, continuum slopes) into high-level contextual representations (e.g., $\rm T_{\rm eff}$). The YOLOv5 backbone is built from ConvBNSiLU and C3 modules. 

Next, The neck serves as a critical intermediary, refining and enhancing the backbone’s hierarchical features through multi-scale aggregation. This hierarchical fusion allows the model to resolve ambiguities in scale or occlusion, while preserving both fine-grained textures and broad structural patterns. Feature Pyramid Networks (FPN) was introduced to deep learning object detectors by \cite{lin2017fpn}. The construction of FPN starts from features of different stages by the backbone. Then upsampling the low-resolution feature map (top feature, form the later stage), and connect it to the high-resolution feature map (bottom featrue, form earlier stage) to form a pyramid. The FPN is a top-down fusion. PAN stands for Path Aggregation Network, which was first proposed by \cite{liu2018pan}. In contrast to FPN that utilizes unidirectional fusion, PAN employs a form of bidirectional fusion. PAN builds on FPN by adding a bottom-up path augmentation, which enables information to flow in both the top-down and bottom-up directions, thus enhance the feature hierarchy with accurate localization signals from lower layers. YOLOv5’s neck does not use standalone FPN or PAN but integrates their principles into a new CSP-PAN architecture. SPPF is also utilized.

Finally, the head produces the ultimate output. For instance, in object detection, the head computes bounding boxes, class probabilities, or other relevant information. This concise architecture ensures effective feature extraction, transformation, and prediction. 

While sharing a similar structure and architecture with YOLOv5, YOLO11 incorporates several improved building blocks. YOLO11 incorporates the C3K2 (Cross Stage Partial with kernel size 2) block, which enhances feature extraction and model efficiency. Additionally, YOLO11 features the C2PSA (Convolutional block with Parallel Spatial Attention), which helps the model focus on relevant parts of the image, thereby improving detection performance. These enhancements, along with optimized components like the SPPF (Spatial Pyramid Pooling - Fast), make YOLO11 more efficient and accurate compared to YOLOv5. More details could be found in their official documentation\footnote{YOLOv5:\url{https://docs.ultralytics.com/models/yolov5/};  YOLO11:\url{https://docs.ultralytics.com/models/yolo11/}}.

\section{supplementary figures}
\label{sec:appendix_figure}

\begin{figure}[h]
\centering
\includegraphics[width=0.7\columnwidth]{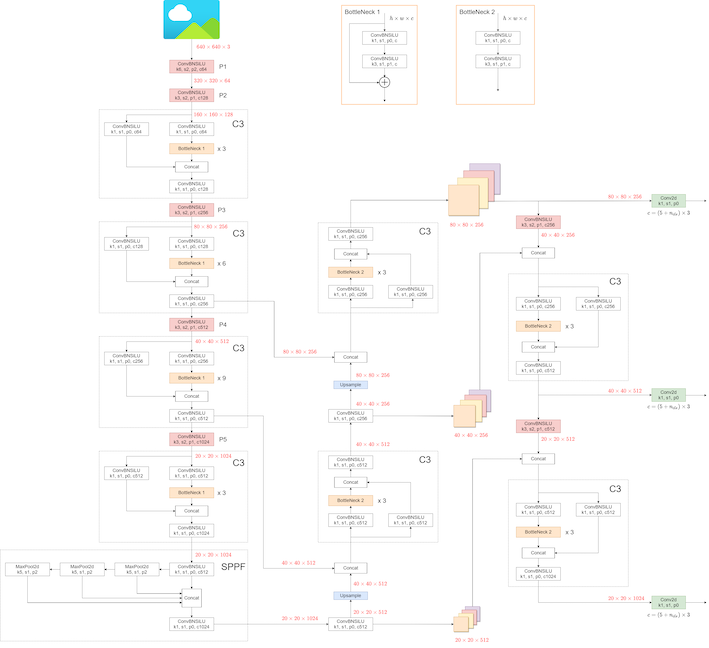}
\centering\caption{the construction of YOLOv5 model from official website}
\label{fig:YOLOv5_cons}
\end{figure}

\begin{figure}[h]%
    \centering
    \begin{subfigure}[\centering Chip 02 (GV), ground truth]{{\includegraphics[width=0.4\columnwidth]{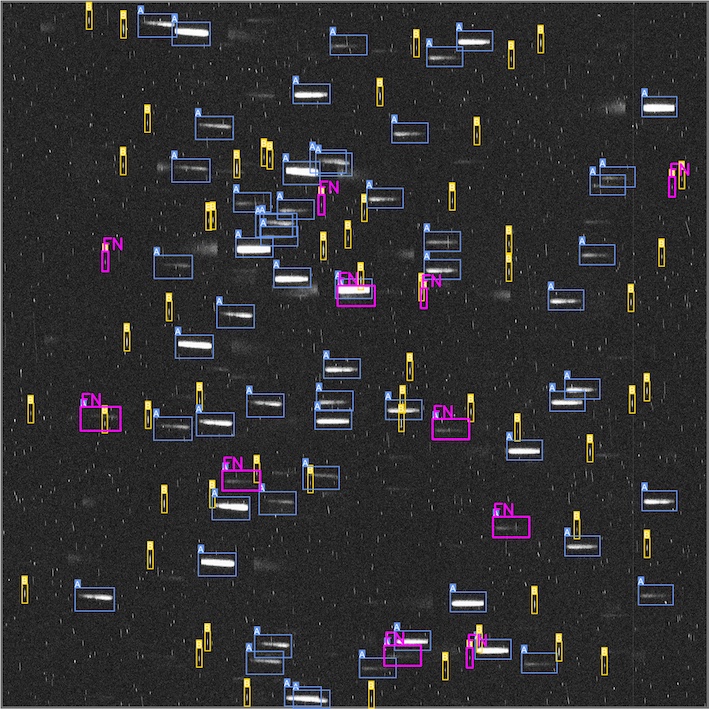}}}%
    \label{}
    \end{subfigure}
    \qquad
    \begin{subfigure}[\centering Chip 02 (GV), predictions]{{\includegraphics[width=0.4\columnwidth]{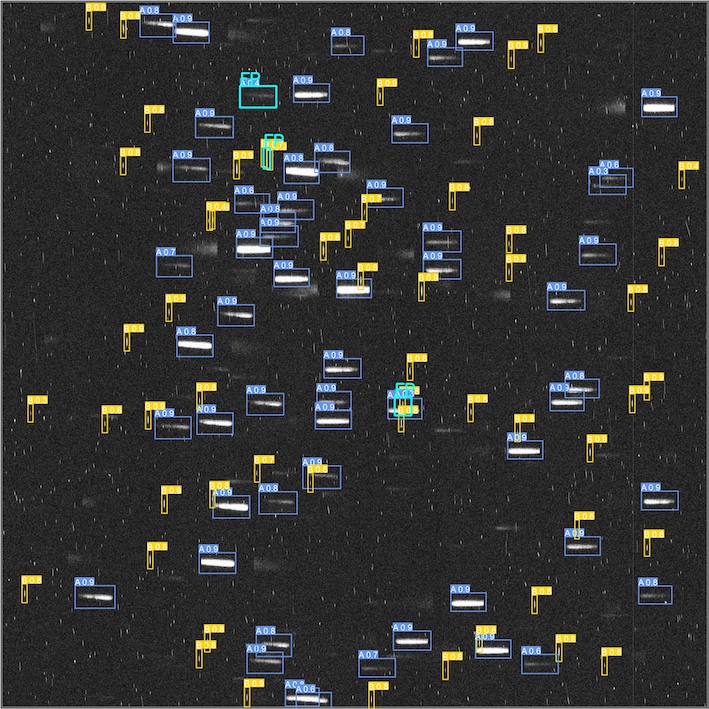}}}%
    \label{}
    \end{subfigure}
    \qquad
    \begin{subfigure}[\centering Chip 03 (GU), ground truth]{{\includegraphics[width=0.4\columnwidth]{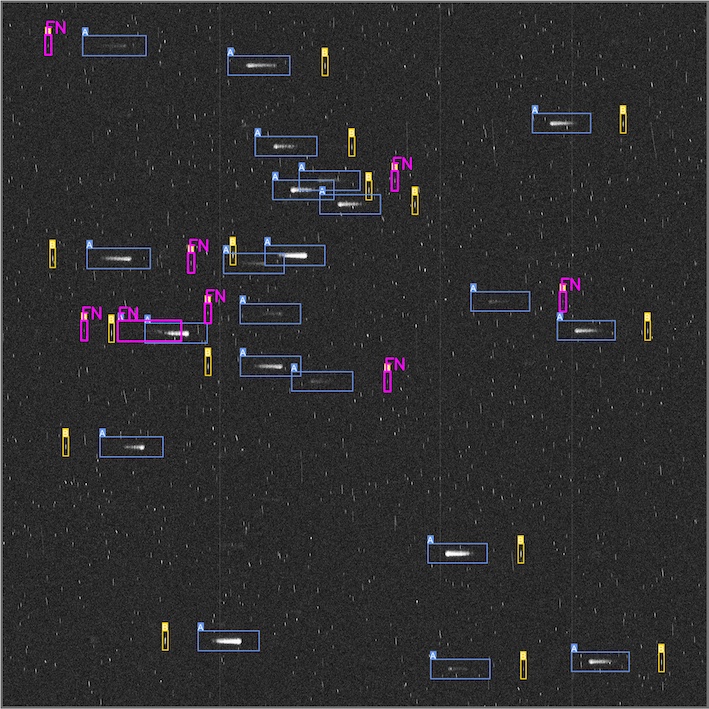}}}%
    \label{}
    \end{subfigure}
    \qquad
    \begin{subfigure}[\centering Chip 03 (GU), predictions]{{\includegraphics[width=0.4\columnwidth]{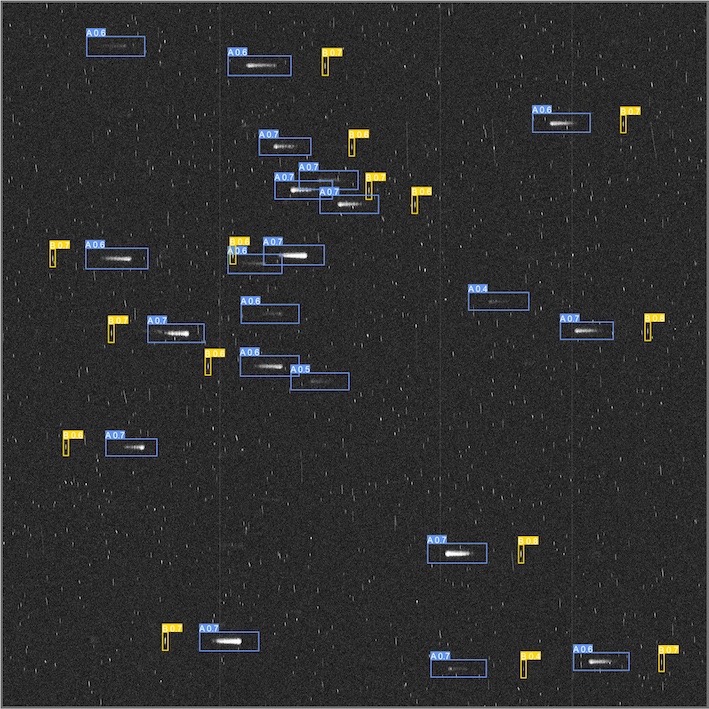}}}%
    \label{}
    \end{subfigure}
    \caption{Similar to Figure \ref{fig:ground_truth_and_pred}, Ground truth (left column, with a magnitude threshold of 22) and predictions (right column, with scores $\ge$ 0.25) from model YOLOv5s. Images are derived from simulations of Chip 02 (GV) and Chip 03 (GU). In these sub-figures, yellow and blue bounding boxes mark Class B and Class A, magenta and cyan boxes highlight false negatives and false positives respectively.
    In GV band, for class A the balance of recall and precision are 0.906 and 0.964; for class B the corresponding values are 0.878 and 0.948. In GU band, for class A the balance of recall and precision are 0.965 and 0.913; for class B the corresponding values are 0.700 and 0.940. }
    \label{fig:ground_truth_and_pred_GVGU}
\end{figure}

\begin{figure}[h]%
    \centering
    \begin{subfigure}[\centering Chip 01 (GI), FP and FN]{{\includegraphics[width=0.3\columnwidth]{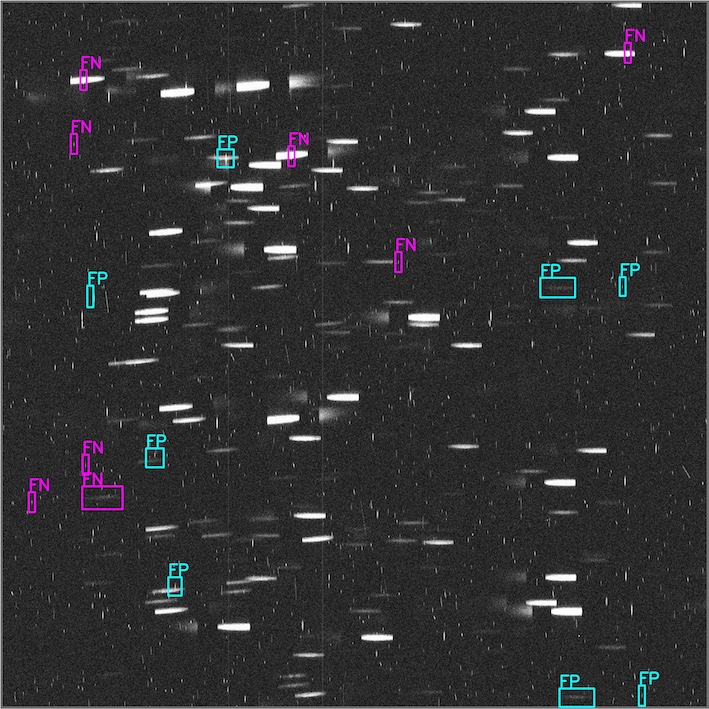} }}%
    \label{}
    \end{subfigure}
   ~~
    \begin{subfigure}[\centering ground truth, $m_{GI0}<22$]{{\includegraphics[width=0.3\columnwidth]{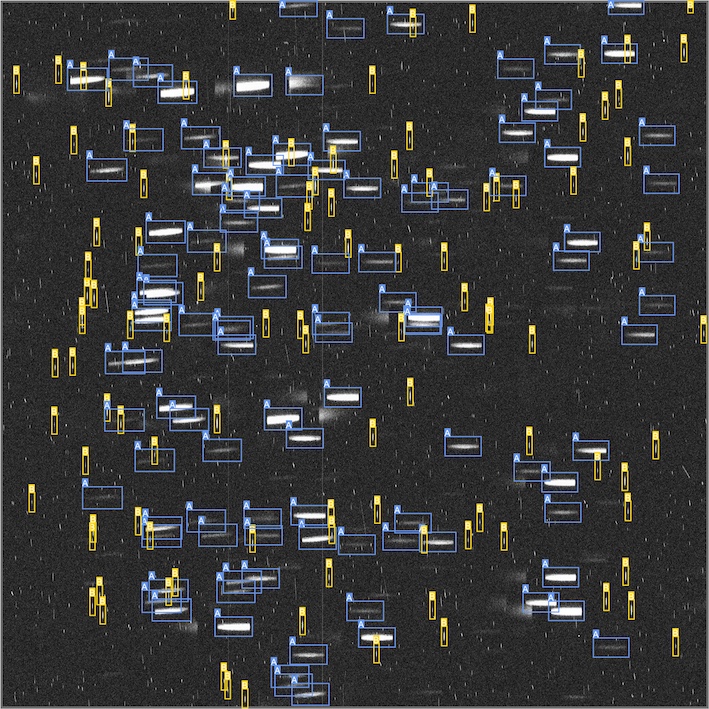}}}%
    \label{}
    \end{subfigure}
   ~~
    \begin{subfigure}[\centering ground truth, $m_{GI0}<22$]{{\includegraphics[width=0.3\columnwidth]{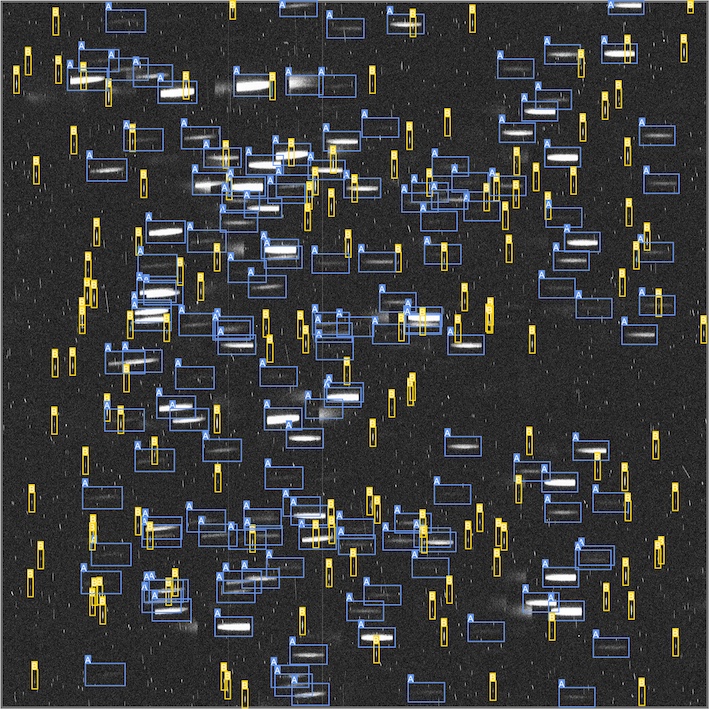} }}%
    \label{}
    \end{subfigure}
    \qquad
    \caption{Visualization of false positives and false negatives. (a) The false positives (FP, shown in cyan) and false negatives (FN, shown in magenta), which are based on the ground truth defined with a magnitude threshold of 22. (b) The ground truth defined with a magnitude threshold of 22. (c) The ground truth defined with a magnitude threshold of 23, which contains more positive samples than sub-figure (b). By analyzing these three figures, we can infer that (1) for the majority of false negatives (missed detections), the misclassification is due to their low brightness; (2)  Many of the "false positives" are actually positives but due to their faintness they are not classified as ground truth.}
    
    \label{fig:fnfp}
\end{figure}

\begin{figure*}[h]%
    \centering
    \begin{subfigure}[\centering Chip 02 (GV), ground truth]{{\includegraphics[width=0.3\columnwidth]{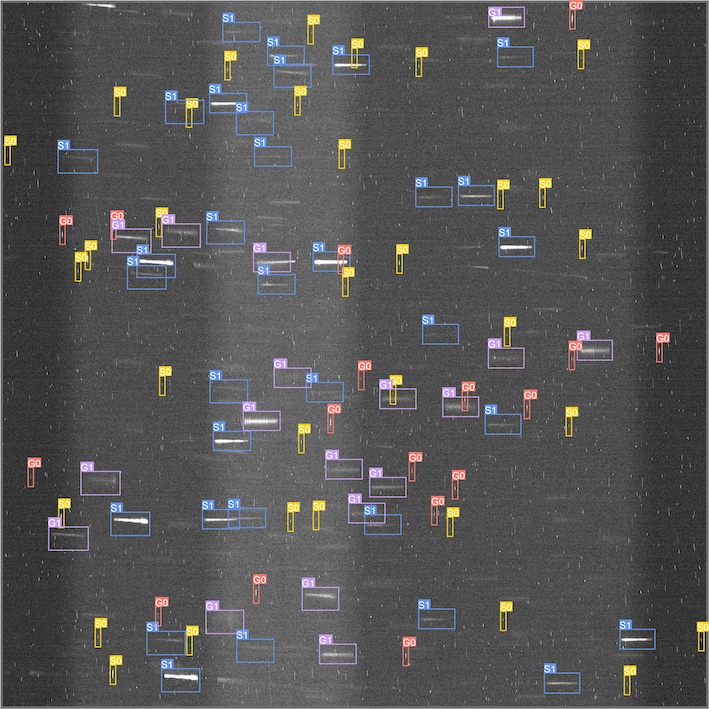} }}%
    \label{fig:YOLO11_sg_gv_a}
    \end{subfigure}
    ~~
    \begin{subfigure}[\centering YOLO11s predictions]{{\includegraphics[width=0.3\columnwidth]{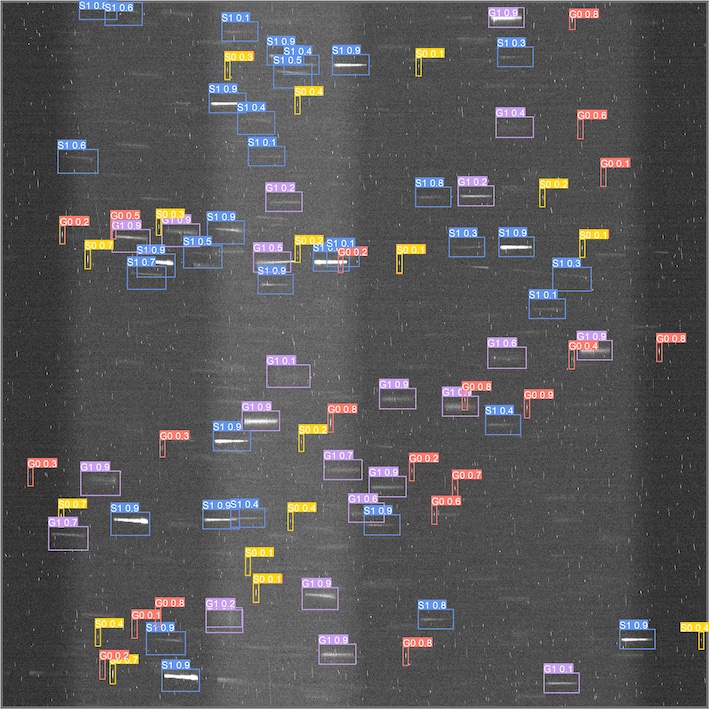}}}%
    \label{fig:YOLO11_sg_gv_b}
    \end{subfigure}
    ~~
    \begin{subfigure}[\centering confusion matrix]{{\includegraphics[width=0.3\columnwidth]{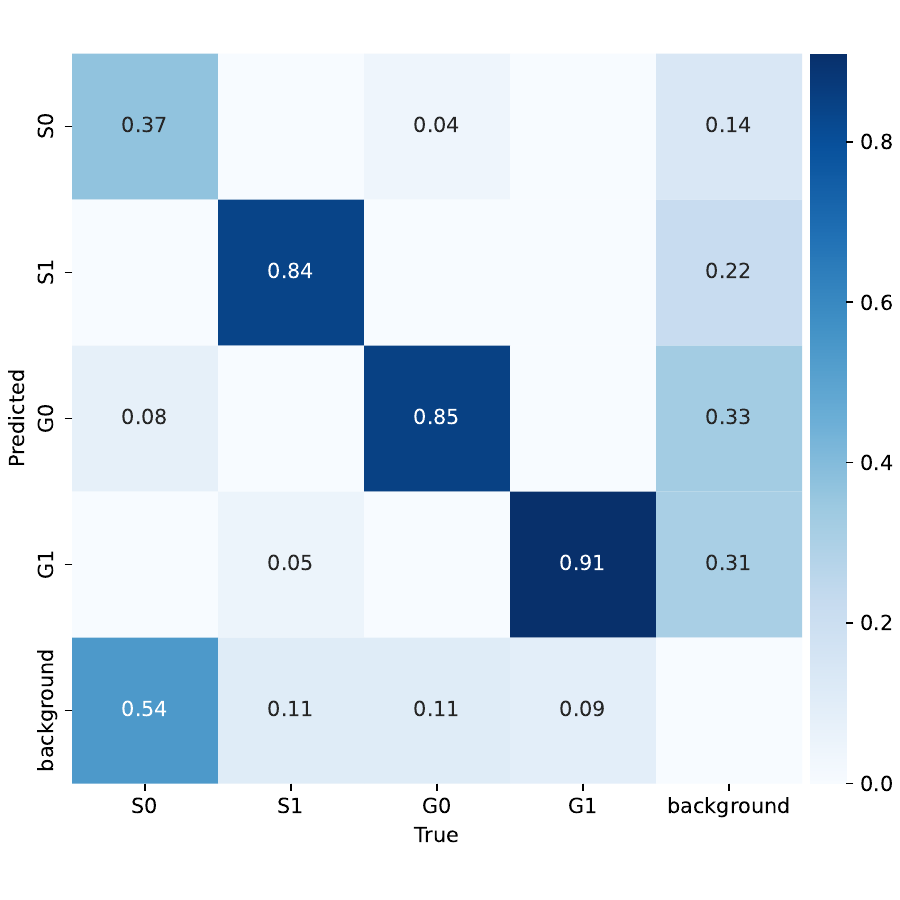}}}%
    \label{fig:YOLO11_sg_gv_c}
     \end{subfigure}

    \caption{The performance of YOLO11 model on detecting stars and galaxies, for Chip 02 (GV). (a) the ground truth, with a magnitude threshold of 23 for stars, 21 for galaxies; (c) the predictions, with scores showing in each box; (b) False Negatives (missed detection) - magenta, False Positives (false detection) - cyan; (d) the confusion matrix of the test set that contains 100 sub-figures like (a). In (d) the diagonal elements shows the recall rates of these four classes.}

    \label{fig:YOLO11s_sg_gv}
\end{figure*}

\begin{figure*}[h]%
    \centering
    \begin{subfigure}[\centering Chip 03 (GU), ground truth]{{\includegraphics[width=0.3\columnwidth]{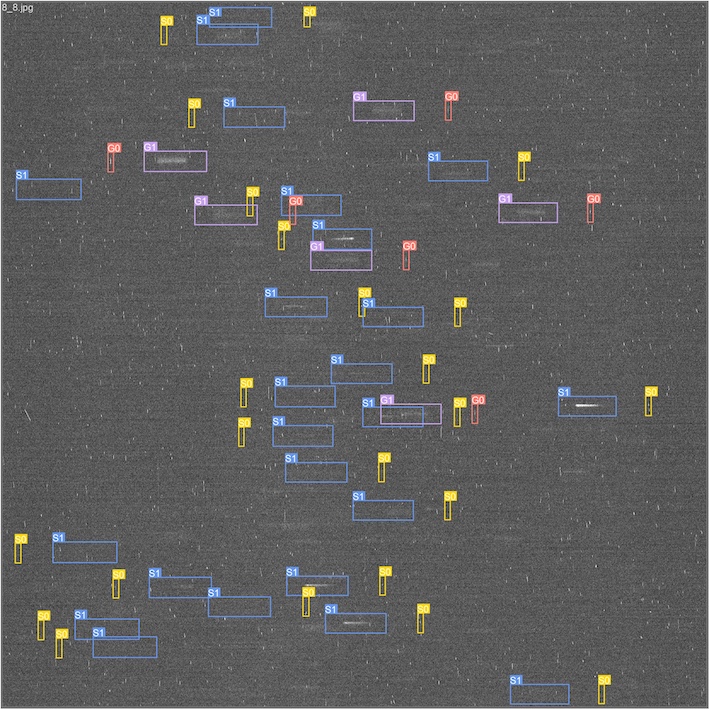} }}%
    \end{subfigure}
    ~~
    \begin{subfigure}[\centering YOLO11s predictions]{{\includegraphics[width=0.3\columnwidth]{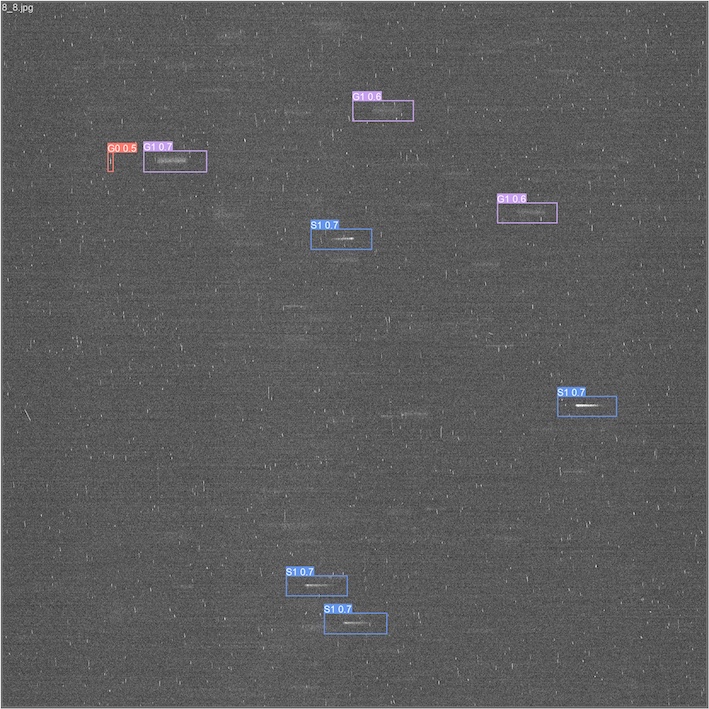}}}%
    \end{subfigure}
    ~~
    \begin{subfigure}[\centering confusion matrix]{{\includegraphics[width=0.3\columnwidth]{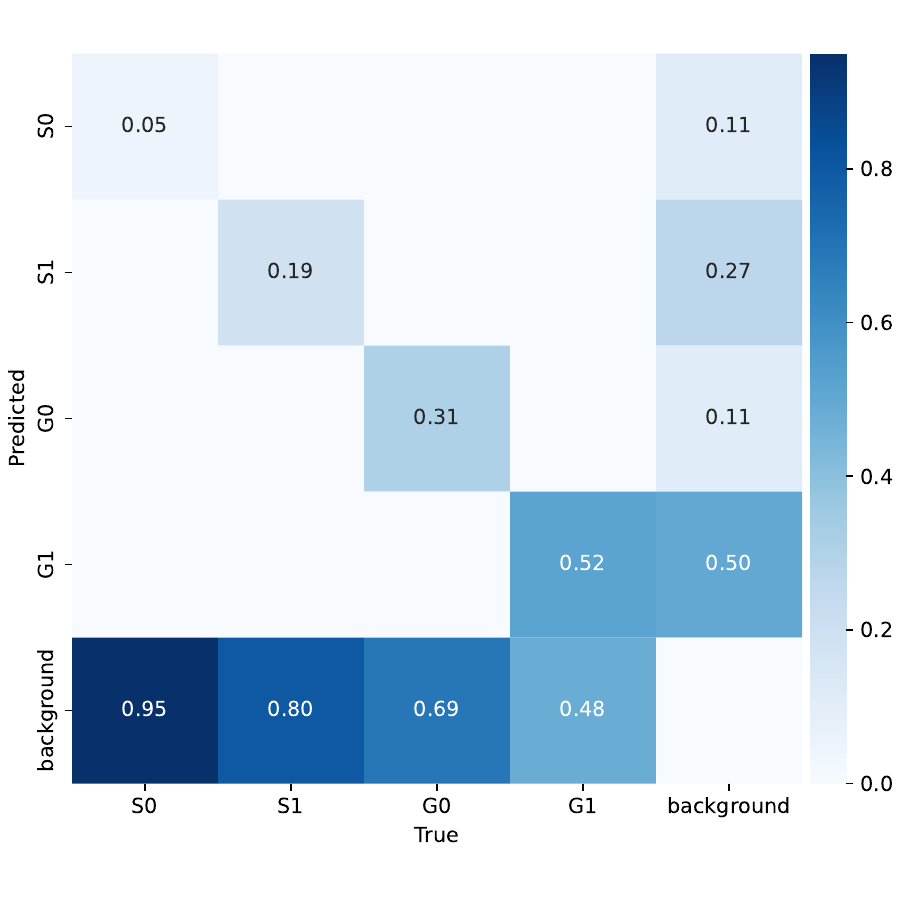}}}%
     \end{subfigure}

    \caption{Similar to Figure \ref{fig:YOLO11s_sg_gv}, but for simulations of Chip 03 (GU). Detection in this band is challenging, potentially due to the low SNR.}

    \label{fig:YOLO11s_sg_gu}
\end{figure*}

\begin{figure*}[h]%
    \centering
    \includegraphics[width=0.8\columnwidth]{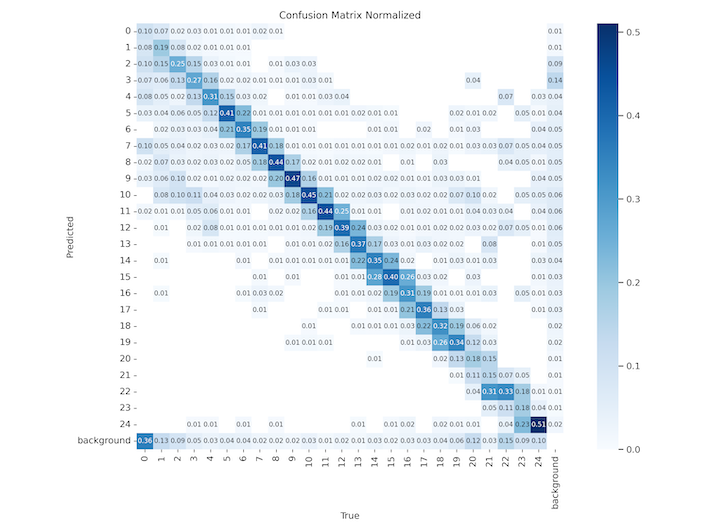}

    \caption{Normalized confusion matrix for galaxy redshift classification. Galaxy redshifts are grouped into 25 bins. The diagonal elements represent correct predictions, showing the model's accuracy within each redshift range.}

\label{fig:YOLO11s_galaxy_redshift_24}
\end{figure*}

\bibliographystyle{aasjournal}
\bibliography{yolo}

\end{document}